\documentclass[onefignum,onetabnum]{siamart190516}

\usepackage{booktabs}  
\usepackage{subcaption} 
\usepackage{balance}            
\usepackage{prftree}
\usepackage{framed}
\usepackage{relsize}
\usepackage{mathtools, cuted}
\usepackage[linesnumbered,ruled,vlined]{algorithm2e}
\usepackage{stfloats}
\usepackage{fullpage}
\usepackage{microtype}
\usepackage{subcaption}

\newtheorem{example}{Example}

\DeclareMathOperator{\SelectNC}{SelectNC}
\DeclareMathOperator{\SelectG}{SelectG}
\DeclareMathOperator{\SelectT}{SelectT}
\DeclareMathOperator{\Select}{Select}
\DeclareMathOperator{\Max}{Max}
\DeclareMathOperator{\Dep}{dep}
\DeclareMathOperator{\Len}{len}
\DeclareMathOperator{\Var}{var}
\DeclareMathOperator{\ComputeTop}{ComputeTop}
\DeclareMathOperator{\Conden}{Conden}

\DeclareMathOperator{\TRes}{TRes}

\DeclareMathOperator{\Saturate}{Saturate}

\usepackage{lipsum}
\usepackage{amsfonts}
\usepackage{graphicx}
\usepackage{epstopdf}
\usepackage[linesnumbered,ruled,vlined]{algorithm2e}
\ifpdf
  \DeclareGraphicsExtensions{.eps,.pdf,.png,.jpg}
\else
  \DeclareGraphicsExtensions{.eps}
\fi

\newsiamremark{remark}{Remark}
\newsiamremark{hypothesis}{Hypothesis}
\crefname{hypothesis}{Hypothesis}{Hypotheses}
\newsiamthm{claim}{Claim}

\title{Querying Guarded Fragments via Resolution}

\author{Sen Zheng\thanks{Department of Computer Science, University of Manchester, UK, \email{sen.zheng@postgrad.manchester.ac.uk}}
\and Renate A. Schmidt\thanks{Department of Computer Science, University of Manchester, UK, \email{renate.schmidt@manchester.ac.uk}}}

\usepackage{amsopn}

\ifpdf
\hypersetup{
  pdftitle={Querying Guarded Fragments via Resolution},
  pdfauthor={Sen Zheng, Renate A. Schmidt}
}
\fi

\begin{document}

\maketitle

\begin{abstract}
Answering Boolean conjunctive queries over the guarded fragment is decidable, however, as yet no practical decision procedure exists. Meanwhile, ordered resolution, as a practically oriented algorithm, is widely used in state-of-art modern theorem provers. In this paper, we devise a resolution decision procedure, which not only proves decidability of querying of the guarded fragment, but is implementable in any `off-the-shelf' resolution theorem prover with modest effort. Further, we extend the procedure to querying the loosely guarded fragment. 

The difficulty in querying a knowledge base of (loosely) guarded clauses is that query clauses are not guarded. We show however there are ways to reformulate query clauses into (loosely) guarded clauses either directly via the separation and splitting rules, or via performing inferences using our top-variable inference system combining with a form of dynamic renaming. Therefore, the problem of querying the (loosely) guarded fragment can be reduced to deciding the (loosely) guarded fragment and possibly irreducible query clauses, i.e., a clause that cannot derive any new conclusion. Meanwhile, our procedure yields a goal-oriented query rewriting algorithm: Before introducing datasets, one can produce a saturated clausal set $\mathcal{S}$ using given BCQs and (loosely) guarded theories. Clauses in $\mathcal{S}$ can be easily transformed first-order queries, therefore query answering over the (loosely) guarded fragment is reduced to evaluating a union of first-order queries over datasets. As far as we know, this is the first practical decision procedure for answering and rewriting Boolean conjunctive queries over the guarded fragment and the loosely guarded fragment.
\end{abstract}



\section{Introduction}
Answering queries over rules and dependencies is at the heart of knowledge representation and database research. In this paper, we are interested in the problem of answering Boolean conjunctive queries over the (loosely) guarded fragment. A \emph{Boolean conjunctive query} (BCQ) is a first-order formula of the form $\exists \overline x \varphi(\overline x)$ where $\varphi$ is a conjunction of atoms containing only constants and variables as arguments. Given a Boolean conjunctive query $q$, a set $\Sigma$ of rules and a database $\mathcal{D}$, the aim is to check whether $\Sigma \cup \mathcal{D} \models q$. Many vital problems, such as query evaluation, query entailment \cite{baget2011decidability} and query containment in database research~\cite{chandra1977querycontainment}, and constraint-satisfaction problems and homomorphism problems in general AI research \cite{vardi2000constraint} can be recast as a BCQ answering problem. We consider the case where rules $\Sigma$ are represented in the guarded fragment \cite{andrka1998bounded} or the loosely guarded fragment \cite{van1997dynamic}. Formulas in the guarded fragment (GF) are first-order formulas, in which the quantification is restricted to the form $\forall \overline x(G \to \varphi)$ and $\exists \overline x(G \land \varphi)$ with the so-called \emph{guard} $G$ containing all free variables of $\varphi$. Satisfiability in many decidable propositional modal logics such as $\mathcal{K}$, $\mathcal{D}$, $\mathcal{S}3$, $\mathcal{S}4$ and $\mathcal{B}$ can be encoded as satisfiability of formulas in GF. The loosely guarded fragment (LGF) further extends GF by allowing multiple guards, allowing, for instance, LGF to express the \emph{until} operator in temporal logic. Both GF and LGF inherit robust decidability, captured by the tree model property \cite{vardi1996robust}, from modal logic \cite{erich1999guards,hadkinson2002loosely}, and have been extensively investigated from a theoretical perspective \cite{gradel1999guarded,andrka1998bounded,erich1999guards}. Practical procedures for deciding satisfiability in GF and LGF have been developed in \cite{hladik2002saga,de2003deciding,ganzinger1999superposition,zheng2020horn}. 

Using conjunctive query to retrieve an answer from GF is undecidable, which is implied by the undecidability of GF with transitivity is undecidable \cite{gradel1999guarded}. E.g., a conjunctive query $q_c$ of the form $q(x,z) \gets \exists y (Axy \land Byx)$ is more expressive than the transitivity axiom, thus answering the query $q_c$ over GF itself is undecidable. However, BCQ answering for GF is \textsc{2ExpTime}-complete \cite{barany2010querying}, and satisfiability checking for the clique-guarded negation fragment, which subsumes both LGF and (negated) BCQs, is also \textsc{2ExpTime}-complete \cite{barany2015negation}. These results mean that BCQ answering for both GF and LGF are decidable, however, as yet there has been insufficient effort to develop practical decision procedures to solve these problems. In this paper, we use resolution to solve BCQ answering and rewriting problems in GF and LGF. Resolution provides a powerful method for developing practical decision procedures as has been shown in~\cite{de2003deciding,ganzinger1998resolution,ganzinger1999superposition,dixon1998resolution,hustadt1999maslov,hustadt1999resolution,hustadt1997evaluating,bachmair1993superposition}. 

%


One of the main challenges in this work is the handling of query formulas, since these formulas, e.g., $\exists xyz (Rxy \land Ryz)$, are beyond (L)GF. By simply negating a BCQ, one can obtain a \emph{query clause}, that is, a clause of only negative literals containing only variables and constants, such as $\lnot Rxy \lor \lnot Ryz$. One can take query clauses as (hyper-)graphs where variables are vertices and literals are (hyper-)edges. Then we use a separation rule \textbf{Sep} \cite{schmidt2000fluted}, which is also referred to as `splitting through new predicate symbol' \cite{kazakov2006Phd}, and the splitting rule~\textbf{Split} \cite{bachmair2001resolution} to cut branches off query clauses. Each `cut branch' follows the guardedness pattern, namely is a \emph{guarded clause}. In general, we found that if a query clause $Q$ is acyclic, one can rewrite~$Q$ into a set of Horn guarded clauses by exhaustively applying separation and splitting to $Q$. That an acyclic BCQ can be equivalently rewritten as a query in the form of guarded clause (hence a bounded hypertree width query) is also reflected in other works \cite{gottlob2003hypertree,barany2015negation}. If a query clause is cyclic, after cutting all branches, one can obtain a new query clause~$Q$ that only consists of variable cycles, i.e., each variable links literals containing different and non-inclusion variable sets. E.g., $\lnot Axy \lor \lnot Byz \lor \lnot Czx$. We use top variable resolution \textbf{TRes} to handle such query clauses, so that by resolving multiple literals in $Q$, the variable cycles are broken. Then we use a dynamic renaming technique \textbf{T-Trans}, to transform a \textbf{TRes}-resolvent into a smaller query clause and a set of (loosely) guarded clauses. We show that only finitely many definers are introduced by \textbf{Sep} and \textbf{T-Trans}, hence, only finite many new clauses are introduced. Top variable resolution \textbf{TRes} is inspired by the `MAXVAR' technique in deciding the loosely guarded fragment \cite{de2003deciding,ganzinger1999superposition}, which is adjusted in \cite{zheng2020horn} to solve BCQs answering problem over the Horn loosely guarded fragment. Interestingly, separation and splitting in query rewriting behaves like GYO-reduction represented in \cite{yu1979tree}, where cyclic queries $q$ \cite{yannakakis1981acyclic} are identified by recursively removing `ears' in the hypergraph of $q$. This means separation and splitting provide an alternative for identifying cyclic queries.

Having a set of (loosely) guarded clauses, another task is building an inference system to reason with these clauses. Existing inference systems for (L)GF are either based on tableau (see \cite{hladik2002saga,hirsch2002tableau}) or resolution (see \cite{de2003deciding,ganzinger1999superposition,zheng2020horn}). We develop an inference system in line with the framework in \cite{bachmair2001resolution}, as it provides a powerful system unifying many different resolution refinement strategies that exist in different forms, e.g., standard resolution, ordered resolution, hyper-resolution and selection-based resolution. It allows us to take advantage of simplification rules and notions of redundancy elimination. We develop our system as an extension of \cite{ganzinger1999superposition,zheng2020horn}, which are the only existing systems that decide (L)GF.

\subsection*{Contributions}
To best of our knowledge, we give the first practical BCQ answering and (goal-oriented) rewriting procedure \textbf{Q-AR} for the guarded fragment and the loosely guarded fragment (with detailed proofs), so that most `off-the-shelf' resolution theorem provers can be used as query answering and rewriting engines. Additionally, we create and exploit reasoning techniques to handle query clauses, and show that an acyclic query clause is expressible using Horn guarded clauses, and a cyclic query clause can be properly handled by i) resolving multiple literals (whenever there exist side premises of that given query) and ii) renaming into a smaller query clause, and be eventually reduced to an acyclic query clause.

\subsection*{Related Work}
Many interesting works had been done on the problem of BCQ answering over restricted forms of guard fragments, see \cite{kikot2012conjunctive,calvanese2007tractable,mora2014kyrie2,rosati2010improving,calautti2015chase,cali2013taming}. The query rewriting procedure of \cite{cali2013taming} is so-called `squid decomposition' to decomposition queries. Squid decomposition aims to rewrite BCQs over Datalog$^{+/-}$ using the chase approach~\cite{abiteboul1995chase}. In a squid decomposition, a query is regarded as a squid-like graph in which branches are `tentacles' and variable cycles are `heads'. Squid decomposition finds ground atoms that are complementary in the squid head, and then use ground unit resolution to eliminate the heads. In contrast, our approach first uses the \textbf{Sep} and \textbf{Split} rules to cut all `tentacles' in queries, and then uses \textbf{TRes} to break cycles in `heads'. Our procedure captures a compact saturated set of non-ground clauses, which avoids the significant overhead of grounding, thus yielding a more efficient query rewriting procedure.


\section{Preliminaries}
\label{sec:pre}
Let \textbf{C}, \textbf{F}, \textbf{P} denote pairwise disjoint discrete sets of \emph{constant symbols} $c$, \emph{function symbols} $f$ and \emph{predicate symbols} $P$, respectively. A \emph{term} is either a variable or a constant or an expression~$f(t_1, \ldots ,t_n)$ where $f$ is a $n$-ary function symbol and $t_1, ... , t_n$ are terms. A \emph{compound term} is a term that is neither a variable nor a constant. A \emph{ground term} is a term containing no variables. An \emph{atom} is an expression $P(t_1, \ldots ,t_n)$, where $P$ is an $n$-ary predicate symbol and $t_1, \ldots , t_n$ are terms. A \emph{literal} is an atom $A$ (a \emph{positive literal}) or a negated atom $\lnot A$ (a \emph{negative literal}). The terms $t_1, \ldots, t_n$ in literal $L = P(t_1, \ldots ,t_n)$ are the \emph{arguments} of $L$. A \emph{first-order clause} is a multiset of literals, presenting a disjunction of literals. An \emph{expression} can be a term, an atom, a literal, or a clause. A \emph{compound-term} literal (clause) is a literal (clause) that contains at least one compound term argument. We use $\Dep(t)$ to denote the depth of a term $t$, formally defined as: if $t$ is a variable or a constant, then $\Dep(t) = 0$; and if $t$ is a compound term $f(u_1, \ldots, u_n)$, then $\Dep(t) = 1 + max(\{\Dep(u_i) \ | \ 1 \leq i \leq n \})$. In a first-order clause $C$, the \emph{length} of $C$ means the number of literals occurring in $C$, denoted as $\Len(C)$, and the \emph{depth} of $C$ means the deepest term depth in~$C$, denoted as $\Dep(C)$. Let $\overline x$, $\mathcal{X}$, $\mathcal{L}$, $\mathcal{C}$ denote a sequence of variables, a set of variables, a set of literals and a set of clauses, respectively. Let $\Var(E)$ be a set of variables in an expression $E$.

As customary, we use the Herbrand universe \cite{abiteboul1995chase} as the domain. The rule set $\Sigma$ denotes a set of first-order formulas and the database $\mathcal{D}$ denotes a set of ground atoms. A \emph{Boolean conjunctive query} (BCQ) $q$ is a first-order formula of the form $\exists \overline x \varphi(\overline x)$ where $\varphi$ is a conjunction of atoms containing only constants and variables. We answer a Boolean conjunctive query $\Sigma \cup D \models q$ by checking the satisfiability of $\Sigma \cup D \cup \lnot q$.


\section{From logic fragments to clausal sets}
\label{sec:clause}
We first provide the formal definitions of GF and LGF, and then transform BCQs and (loosely) guarded formulas into suitable sets of clauses.
\begin{definition}[(Loosely) Guarded Fragment]
\label{def:GF}
Without equality and function symbols, the \emph{(loosely) guarded fragment ((L)GF)} is a class of first-order formulas, inductively defined as follows:
\begin{enumerate}
\item $\top$ and $\bot$ belong to (L)GF.
\item If $A$ is an atom, then $A$ belongs to (L)GF.
\item (L)GF is closed under Boolean combinations.
\item Let $F$ belong to $GF$ and $G$ be an atom. Then $\forall \overline x (G \to F)$ and $\exists \overline x (G \land F)$ belong to $GF$ if all free variables of $F$ are among variables of $G$. $G$ is referred to as \emph{guard}.
\item Let $F$ belong to $LGF$ and $\Delta= G_1 \land \ldots \land G_n$ be a conjunction of atoms. Then $\forall \overline x (\Delta \to F)$ and $\exists \overline x (\Delta \land F)$ belong to $LGF$ if
\begin{enumerate}
\item all free variables of $F$ are among the variables of $\Delta$, and 
\item for each variable $x$ in $\overline x$ and each variable $y$ in $\Delta$ distinct from $x$, $x$ and $y$ co-occur in a literal in $\Delta$. $\Delta$ are referred to as \emph{loose guards}.
\end{enumerate}
\end{enumerate}   
\end{definition}

\begin{example}
\label{example:GF}
Some (counter-)examples of guarded formulas:
\begin{gather*}
F_1 = Ax \quad  F_2 = \forall x Ax \quad  F_3 = \forall x(Axy \to Bxy) \quad  F_4 = \forall x (Axy \to \exists yByz) \quad  F_5 = \forall x (Axy \to \bot)\\ 
F_6 = \exists x (Axy \land \forall z (Bxz \to \exists u Czu)) \qquad F_7 = \forall x (Px \to \exists y (Rxy \land \forall z (Ryz \to Pz))))
\end{gather*}
Formulas $F_1, F_3, F_5, F_6$ and $F_7$ are guarded formulas. Formulas $F_2$ and $F_4$ are not guarded formulas since guards are missing. Formula $F_7$ is the standard first-order translation of the modal axiom $P \to \lozenge \square P$ and the description logic axiom $P \sqsubseteq \exists R. \forall R. P$.
\end{example}

LGF strictly extends GF by allowing a conjunction of loose guards in the guard position. E.g., the first-order translation of a temporal logic formula \text{$P$ \emph{until} $Q$} is a loosely guarded formula $\exists y (Rxy \land Qy \land \forall z((Rxz \land Rzy)\to Pz)))$, but not guarded. The transitivity axiom $\forall xyz ((Rxy \land Ryz) \to Rxz)$ is neither guarded nor loosely guarded. Satisfiability checking of (L)GF with transitivity is indeed undecidable \cite{erich1999guards}.

\subsection*{Clausal normal form transformation \textbf{Q-Trans}}
We use \textbf{Q-Trans} to denote our \emph{clausal normal form transformation} for (loosely) guarded formulas and BCQs. \textbf{Q-Trans} simply negates a BCQ to obtain a \emph{query clause}, which is denoted as $Q$. \textbf{Q-Trans} transforms (loosely) guarded formulas as the ones in \cite{de2003deciding,ganzinger1999superposition,zheng2020horn} (see details in \textbf{Appendix \ref{appen:clause}}). E.g., $F_6 = \exists x (Axy \land \forall z (Bxz \to \exists u Czu))$ in \textsc{Example \ref{example:GF}} is transformed into a clausal set $\{Aab, d_{\forall}^{1}a, \lnot d_{\forall}^{1}x \lor \lnot Bxz \lor C(z,fxz)\}$ where $a$ and $b$ are Skolem constants, $f$ is a Skolem function and $d_{\forall}^{1}$ is a new predicate symbol, namely definer, introduced by formula renaming.

We need the notions of \emph{flatness}, \emph{simpleness} and \emph{covering} to properly define (loosely) guarded clauses and query clauses in \textsc{Definitions} \ref{def:query}--\ref{def:gc} below. A literal $L$ is \emph{flat} if each argument in $L$ is either a constant or a variable. A literal $L$ is \emph{simple} \cite{ganzinger1999superposition} if each argument in $L$ is either a variable or a constant or a compound term $f(u_1, \ldots, u_n)$ where each $u_i$ is either a variable or a constant. A clause $C$ is called \emph{simple} (\emph{flat}) if all literals in $C$ are simple (flat). A clause $C$ is \emph{covering} if each compound term $t$ in $C$ satisfies $\Var(t)=\Var(C)$. 


\begin{definition}
\label{def:query}
A \emph{query clause} is a flat first-order clause containing only negative literals.   
\end{definition}
\begin{definition}[(Loosely) Guarded Clause]
\label{def:gc}
\setlength\itemsep{0.05em}
A \emph{(loosely) guarded clause} $C$ is an equality-free, simple and covering first-order clause satisfying the following conditions:
\begin{enumerate}
\item[\emph{1.}] $C$ is either ground, or
\item[\emph{2a.}] if $C$ is a guarded clause, then $C$ contains a negative flat literal $\lnot G$ satisfying $\Var(C) = \Var(G)$. ($G$ is referred to as \emph{guard}), and 
\item[\emph{2b.}] if $C$ is a loosely guarded clause, then $C$ contains a set  $\mathcal{G} = \{\lnot G_1, \ldots, \lnot G_n\}$ of negative flat literals such that each pair of variables in $C$ co-occurs in a literal in $\mathcal{G}$. $\mathcal{G}$ is referred to as \emph{loose guards}.
\end{enumerate}   
\end{definition}

A guarded clause is not necessarily a query clause, and vice versa. E.g., $\lnot Axy \lor \lnot Bx$ is guarded and a query clause. $\lnot Axy \lor B(fxy)$ is guarded but not a query clause, and $\lnot Axy \lor \lnot Byz$ is a query clause, but not guarded. The notion of loosely guarded clauses strictly extends that of guard clauses, since one can reduce the number of loose guards (Condition 2b in \textsc{Definition} \ref{def:gc}) to one to obtain a guarded clause. 
 


\section{Top-variable inference system \emph{T-Inf}}
\label{sec:tinf}
The basis for our querying procedure is the top-variable inference system, first introduced in \cite{zheng2020horn} based on \cite{ganzinger1999superposition}. We enhance the system with the splitting rule, the separation rule, and show that the system is sound and refutationally complete. A crucial difference to system in \cite{zheng2020horn} is that we consider simple forms of clauses while the system in \cite{zheng2020horn} allows clauses with nested ground compound terms, we give simplified proofs and use a simpler menthod (using term depth, rather than variable depth) to compute top variables. \emph{T-Inf} uses ordered resolution, with selection and the non-standard top-variable resolution technique. Clauses are ordered using admissible orderings $\succ$, so that a clause is only allowed to derive smaller (w.r.t. $\succ$) clauses. The notions of admissible orderings and selection functions are given in \textbf{Appendix \ref{appen:tinf}}.

Before we define the specific ordering and selection refinement that gives us a decision procedure for our class of clauses, we define the \emph{T-Inf} calculus.

\subsection*{The calculus} 
The top variable inference system \emph{T-Inf} contains the following rules \textbf{Deduct}, \textbf{Fact}, \textbf{Res}, \textbf{Conden} and \textbf{Delete}. New conclusions are derived using
\begin{align*}
\prftree[r,l]{\qquad \qquad \qquad \ \ 
\mbox{\vbox{\noindent if $C$ is either a conclusion of \textbf{Res}, \textbf{TRes} and \textbf{Fact} of clauses in $N$.}}}
{\textbf{Deduct}: \ }
  {N}
  {N \cup \{C\}} 
\end{align*}
\begin{displaymath}
\prftree[r,l]{\qquad \qquad \qquad \ 
\mbox{\vbox{\noindent if i) no literal is selected in $C$, ii) $A_1$ is $\succ$-maximal with respect to $C$. \\ $\sigma$ is an mgu of $A_1$ and $A_2$.}}}
{\textbf{Fact}: \ }
  {C \lor A_1 \lor A_2}
  {(C \lor A_1)\sigma}
\end{displaymath}
\begin{displaymath}
\prftree[r,l]{\qquad \qquad
\mbox{\vbox{\noindent
 if 1) either $\lnot A$ is selected, or nothing is selected in $\lnot A \lor D$ and $\lnot A$ is \\ the maximal literal in $\lnot A \lor D$, ii) $B$ is strictly $\succ$-maximal with respect \\ to $D_1$. $\sigma$ is an mgu of $A$ and $B$, and premises are variable-disjoint.
 			}
 	}
 }{\textbf{Res}: \ }
  {B \lor D_1}
  { }
  {\lnot A \lor D}
  {(D_1 \lor D)\sigma}
\end{displaymath}
For decidability, the following two rules are specified. 
\begin{displaymath}
 \prftree[r,l]{\qquad \qquad \quad 
 \mbox{\vbox{\noindent
 if $\Conden(C)$ is a proper subclause and an instance of $C$.}}}{\textbf{Conden}: \ }
  {C}
  {\Conden(C)} 
\end{displaymath}
\begin{displaymath}
 \prftree[r,l]{\qquad \qquad \qquad \quad
 \mbox{\vbox{\noindent
 if $C$ is a tautology, or $N$ contains a variant of $C$.}}}{\textbf{Delete}: \ }
  {N \cup \{C\}}
  {N}
\end{displaymath}

Compared to the top variable inference system in \cite{zheng2020horn}, this system contains also the splitting rule \cite{bachmair2001resolution}, the separation rule \cite{schmidt2000fluted}, and a simpler form of top variable resolution rule, specially devised for resolution computations of flat (loosely) guarded clauses and query clauses. These rules are defined next.

We say a clause $C$ is \emph{separable} into two subclauses $D_1$ and $D_2$ if i) $D_1$ and $D_2$ are non-empty, ii) $C$ can be partitioned into $D_1$ and $D_2$ such that $\Var(D_1) \not \subseteq \Var(D_2)$ and $\Var(D_2) \not \subseteq \Var(D_1)$. A clause is \emph{indecomposable} if it is not separable into two variable-disjoint clauses. Then the splitting rule is
\begin{displaymath}
 \prftree[r,l]{\qquad \qquad \quad \ 
 \mbox{\vbox{\noindent
 if $C$ and $D$ are non-empty and variable-disjoint.}}}
 {\textbf{Split}: \ }
  {N \cup \{C \lor D\}}
  {N \cup \{C\} \ | \ N \cup \{D\}}
\end{displaymath}
\textbf{Split} is not applicable to indecomposable clauses. We partition a non-splittable, but separable clause using
\begin{displaymath}
 \prftree[r,l]{ \quad \ 
 \mbox{\vbox{\noindent
 if i) $C \lor D$ is separable into $C$ and $D$, ii) $\overline x = \Var(C) \cap \Var(D)$, \\ iii) $d_s$ is a fresh predicate symbol.}}}{\textbf{Sep}: \ }
  {N \cup \{C \lor D\}}
  {N \cup \{\lnot d_s(\overline x) \lor C, \ d_s(\overline x) \lor D\}}
\end{displaymath}
This is a replacement rule in which $C \lor D$ is replaced by $\lnot d_s(\overline x) \lor C$ and $d_s(\overline x) \lor D$.  We require that the definer symbols introduced by the \textbf{Sep} are smaller than other predicate symbols in the premises. This ensures that \textbf{Sep} is a simplification rule. Top variable based resolvents are computed using the \textbf{TRes} rule:
\begin{align*}
 \prftree[l]{\textbf{TRes}: \quad}
  {B_1 \lor D_1 \ \ \ldots \ \ B_m \lor D_m \ \ \ldots \ \ B_n \lor D_n}
  { }
  { }
  { }
  {\lnot A_1 \lor \ldots \lor \lnot A_m \lor \ldots \lor \lnot A_n \lor D}
  {(D_1 \lor \ldots \lor D_m \lor \lnot A_{m+1} \lor \ldots \lor \lnot A_{n} \lor D)\sigma}
\end{align*}
if i) there exists an mgu $\sigma^\prime$ such that $B_i\sigma^\prime = A_i\sigma^\prime$ for each $i$ such that $1 \leq i \leq n$, making $\lnot A_1 \lor \ldots \lor \lnot A_m$ top-variable literals (see $\ComputeTop$ in \textbf{Resolution Refinement}) and are selected, and $D$ is positive, ii) no literal is selected in $D_1, \ldots, D_n$ and $B_1, \ldots, B_n$ are strictly $\succ$-maximal with respect to $D_1, \ldots, D_n$, respectively. $\sigma$ is an mgu such that $B_i\sigma = A_i\sigma$ for all $i$ such that $1 \leq i \leq m$, and premises are variable-disjoint.


\textbf{TRes} is devised to avoid term depth increase in the resolvents by carefully picking proper side premises~(using top-variable literals). \textbf{TRes} avoids term depth increase when performing inferences on (loosely) guarded clauses and query clauses (\textsc{Lemma \ref{lem:tres_matching}}). Given a main premise $C$ and a set of side premises $\{C_1, \ldots, C_n\}$, if one can perform an inference on $C$ and $\{C_1, \ldots, C_n\}$ such that all negative literals in $C$ are selected (using `maximal selection'), then one can pick any subset $\mathcal{S}$ of $\{C_1, \ldots, C_n\}$ and preform an `partial inference' on $C$ and $\mathcal{S}$. This partial inference makes the `maximal selection resolution' among $\mathcal{S}$ and $\{C_1, \ldots, C_n\}$ redundant (\textsc{Propositions} \ref{prop:bg_redun}--\ref{pro:pres}). \textbf{TRes} is an instance of such `partial inference', which is a key to maintaining refutational completeness. The notion of redundancy is given in \textbf{Appendix \ref{appen:tinf}}.

\subsection*{Ordering and selection refinement} 
We use \emph{T-Refine} to denote our refinements to guide the inference steps. As orderings we use a lexicographic path ordering $\succ_{lpo}$ \cite{dershowitz1982ordering} with a precedence $f > c > p$ over symbols $f \in \emph{\textbf{F}}$, $c \in \emph{\textbf{C}}$ and $p \in \emph{\textbf{P}}$. The selection over negative literals are handled by three selection functions, namely $\SelectNC$, $\SelectG$ and $\SelectT$. \textbf{Algorithm \ref{algorithm:refine}} describes how the ordering $\succ_{lpo}$ and the selection functions are imposed on clauses ($\succ_{lpo}$ can be replaced by any admissible ordering with the restrictions on the precedence as we use above). 

\begin{algorithm}[h]
\DontPrintSemicolon
 \KwIn{A query clause or a (loosely) guarded clause $C$}
 \KwOut{Eligible literals in $C$}
 \lIf{$C$ is ground}{
  \Return $\Max(C)$
  \tcp*[f]{Apply \textbf{Res}, \textbf{TRes} or \textbf{Fact}}
 }
 \lElseIf{$C$ has negative compound terms}{
  \Return $\SelectNC(C)$
  \tcp*[f]{Apply \textbf{Res}}
 }
  \lElseIf{$C$ has positive compound terms}{
  \Return $\Max(C)$
  \tcp*[f]{Apply \textbf{Res}, \textbf{TRes} or \textbf{Fact}}
 }
  \lElseIf{$C$ is a guarded clause}{
  \Return $\SelectG(C)$
  \tcp*[f]{Apply \textbf{Res}}
 }
 \lElse{
  \Return $\SelectT(C)$
  {\tcp*[f]{Apply \textbf{TRes}}}}
  
  \;
 
  \SetKwFunction{FMain}{$\SelectT$}
  \SetKwProg{Fn}{Function}{:}{}
   \Fn{\FMain{$C$}}{
 Select all negative literals $\mathcal{L}$ in $C$
 {\tcp*[f]{To find mgu in `maximal selection resolution'}}
 
 Find side premises $C_1, \ldots, C_n$ of $C$
 {\tcp*[f]{To satisfy Condition i) in \textbf{TRes}}}
 
 \lIf{$C_1, \ldots, C_n$ exist}{\Return $\ComputeTop(C_1, \ldots, C_n, C)$
 {\tcp*[f]{Top-variable literals}}
 }
 \lElse{
 \Return $\mathcal{L}$
 {\tcp*[f]{Select $\mathcal{L}$ for $C$ has no sufficient side premises}
 }}}
 
\caption{Computing eligible literals using \emph{T-Refine}}
\label{algorithm:refine}
\end{algorithm}

Given a clause $C$, \textbf{Algorithm \ref{algorithm:refine}} determines eligible literals in $C$ in two ways: either the (strictly) $\succ_{lpo}$-maximal literals in $C$ are eligible, denoted as $\Max(C)$; or selected literals in $C$ are eligible, denoted as i) $\SelectNC$ selects one of negative compound-term literals, ii) $\SelectG$ selects one of the guards in a clause, iii) $\SelectT$ selects literals based on \emph{top variables}, as described in Lines 7--11 in \textbf{Algorithm \ref{algorithm:refine}}. \emph{Top-variable literals} in a clause $C$ are computed using $\ComputeTop(C_1, \ldots, C_n, C)$ by the following steps:



%

\begin{enumerate}
	\item Without producing or adding the resolvent, compute an mgu $\sigma^\prime$ among $C_1 = B_1 \lor D_1, \ldots, C_n = B_n \lor D_n$ and $C = \lnot A_1 \lor \ldots \lor \lnot A_n \lor D$ such that $B_i\sigma^\prime = A_i\sigma^\prime$ for each $i$ such that $1 \leq i \leq n$.
	\item Compute the variable order $>_v$ and $=_v$ over the variables of $\lnot A_1 \lor \ldots \lor \lnot A_n$. By definition $x >_v y$ and $x =_v y$, if $\Dep(x\sigma^\prime) > \Dep(y\sigma^\prime)$ and $\Dep(x\sigma^\prime) = \Dep(y\sigma^\prime)$, respectively.
	\item Based on $>_v$ and $=_v$, identify the maximal variables in $\lnot A_1 \lor \ldots \lor \lnot A_n$, denoted as the \emph{top variables}. Literals $\lnot A_1, \ldots, \lnot A_m$ are \emph{top-variable literals} in $\lnot A_1, \ldots, \lnot A_n$ if each literal in $\lnot A_1, \ldots, \lnot A_m$ contains at least one of the top variables. 
\end{enumerate}
\textbf{Algorithm \ref{algorithm:refine}} is adequate for computing eligible literals for any input clause (\textsc{lemmas \ref{lem:algo2_eligible}--\ref{lem:algo1_eligible}}). $\SelectNC$ and $\SelectG$ are standard selection functions that picks a fixed set of negative literals in a clause. However, $\SelectT$ selects negative literals in a clause $C$ depending on the combination with side premises in an application of the \textbf{TRes} rule. Hence, we justify $\SelectT$ by showing that \textbf{TRes} is compatible with the framework in \cite{bachmair2001resolution}. Note that in the application of \emph{T-Inf}, one can use \emph{a-priori checking}, to avoid overheads caused by \emph{a-posteriori checking} (see details in \textbf{Appendix \ref{appen:tinf}}).

\subsection*{Soundness and Refutational Completeness}
We give detailed proofs in \textbf{Appendix \ref{appen:tinf}}.

\begin{theorem}
\emph{T-Inf} is a sound and refutational complete inference system.
\end{theorem}

\section{\emph{T-Inf} decides guarded clauses and loosely guarded clauses}
\label{sec:decide_lgc}
Given a set of (loosely) guarded clauses, we show that \emph{T-Inf} only derives bounded depth and length (loosely) guarded clauses, hence decides these clausal class. 

Considering the behaviour of rules in \emph{T-Inf} on (loosely) guarded clauses, \textbf{Split} and \textbf{Sep} are not applicable to (loosely) guarded clauses since these clauses are not separable. This is because given a guarded clause $C$, a guard contains all variables of $C$, hence one cannot partition that clause. If $C$ is loosely guarded, loose guards $\mathcal{G}$ contains all variables of $C$, and $\mathcal{G}$ cannot be partitioned due to the variable co-occurrence property (Condition 2b in \textsc{Definition} \ref{def:gc}). Hence, only the rules \textbf{Fact}, \textbf{Conden}, \textbf{Res} and \textbf{TRes} derive new conclusions from given (loosely) guarded clauses.

Applying \emph{T-Inf} to (loosely) guarded clauses derives only (loosely) guarded clauses (\textsc{Lemmas \ref{lem:conden}--\ref{lem:tres_guard}}). Since (loosely) guarded clauses are simple clauses, no term depth increase occurs in conclusions of \emph{T-Inf} inferences. The length of these conclusions is also bounded (\textsc{lemma \ref{lem:bounded_width}}). Therefore

\begin{theorem}
\label{thm:decide}
\emph{T-Inf} decides guarded clauses and loosely guarded clauses.
\end{theorem}

\textsc{example \ref{example:unLGC}} in the \textbf{Appendix \ref{appen:decide_lgc}} shows how \emph{T-Inf} decides an unsatisfiable loosely guarded clause set, and how the refinements help avoid unnecessary inferences. 

\section{Handling query clauses}
\label{sec:handle_query}
Now we describe the \textbf{Q-AR} calculus which provide the basis for our the query answering and rewriting procedure. \textbf{Q-AR} consists of \textbf{Conden}, \textbf{Split}, \textbf{Sep}, \textbf{TRes} and a form of dynamic renaming \textbf{T-Trans} to handle query clauses. 

\textbf{Conden} and \textbf{Split} immediately replace a query clause by smaller query clauses (with fewer literals) (\textsc{lemma \ref{lem:con_split}}). E.g., \textbf{Conden} replaces $\lnot A(x_1, x_2) \lor \lnot A(a, x_3)$ by its condensation $\lnot A(a, x_3)$, and \textbf{Split} partitions a decomposable query clause $\lnot A(x_1, x_2) \lor \lnot B(y_1, y_2)$ into the indecomposable subclauses $\lnot A(x_1, x_2)$ and $\lnot B(y_1, y_2)$, which are query clauses. Since these rules replace a query clauses with smaller (w.r.t. $\succ_{lpo}$) equisatisfiable query clauses, we use \textbf{Conden} and \textbf{Split} whenever possible. Thus, we can limit our attention to indecomposable and condensed query clauses.


\subsection*{\textbf{Sep} simplifies query clauses into guarded clauses (and possibly chained-only query clauses)} 
\begin{figure}[h]
\captionsetup{singlelinecheck=off}
\begin{minipage}[c]{0.45\textwidth}
  \includegraphics[width=\textwidth]{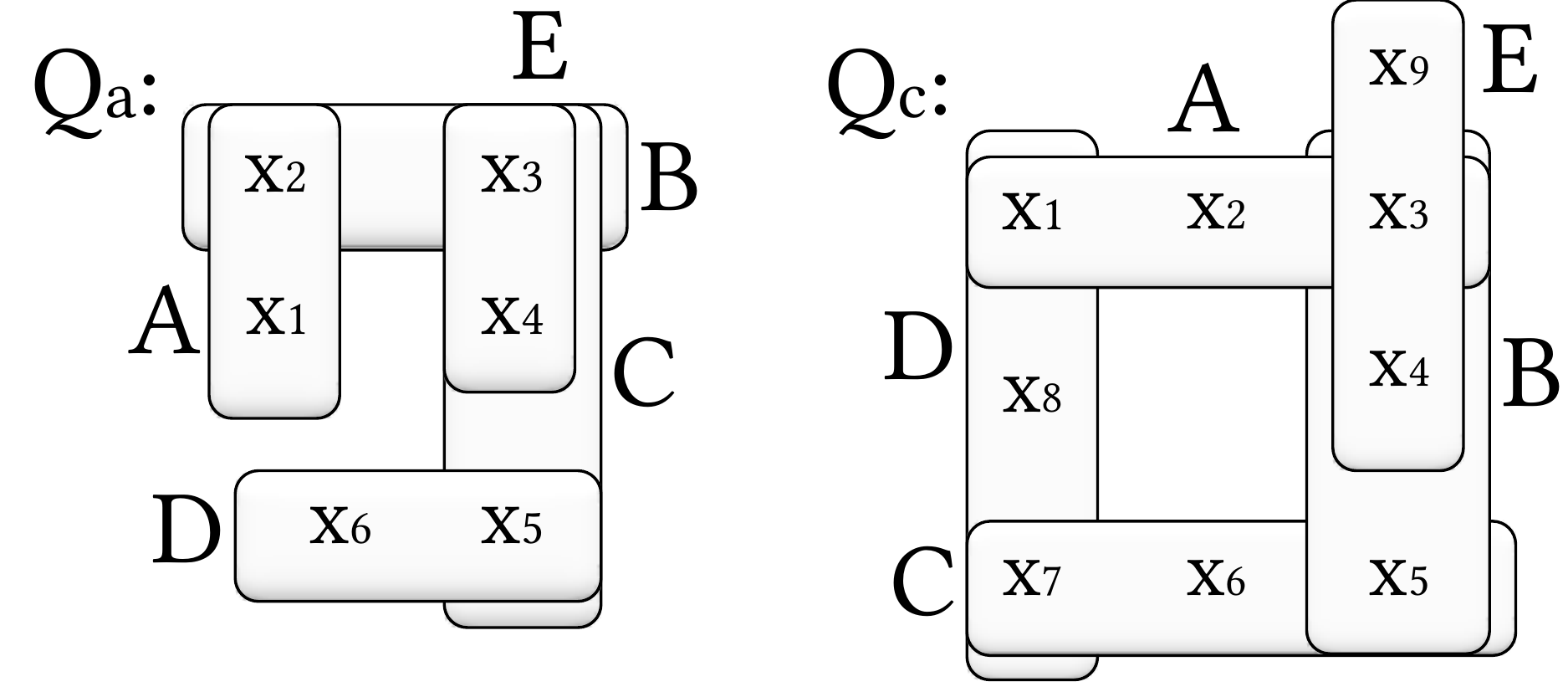}
 \end{minipage}\hfill
 \begin{minipage}[c]{0.52\textwidth}
	\caption[l]{
	Hypergraphs of $Q_a$ and $Q_c$. 
	
	$\{\lnot A_1(x_1,x_2), \lnot B(x_2,x_3), \lnot C(x_3, x_4, x_5), \lnot D(x_5, x_6)\}$ are surface literals in $Q_a$, hence in $Q_a$,
	\begin{itemize}
	\item $x_2, x_3, x_5$ are chained variables, and
	\item $x_1, x_4, x_6$ are isolated variables.
	\end{itemize}
	All literals in $Q_c$ are surface literals, hence in $Q_c$,
	\begin{itemize}
	\item $x_1, x_3, x_4, x_5, x_7$ are chained variables, and
	\item $x_2, x_6, x_8, x_9$ are isolated variables.
	\end{itemize}
	}
	\label{fig:queries}
 \end{minipage}
\end{figure}

First let us analyse variables occurring in query clauses. We use the notion of \emph{surface literal} to divide variables in a query clause into two kinds, namely \emph{chained variables} and \emph{isolated variables}. $L$ is a \emph{surface literal} in a query clause $Q$ if for any $L^\prime$ in $Q$ that is distinct from $L$, \text{$\Var(L) \not \subset \Var(L^\prime)$}. Let the surface literals in a query clause $Q$ be $L_1, \ldots, L_n$ ($n > 1$). Then the \emph{chained variables} in $Q$ are the variables in $\underset{i,j \in n}{\bigcup} \Var(L_i) \cap \Var(L_j)$ whenever $\Var(L_i) \not = \Var(L_j)$, i.e., variables that link distinct surface literals containing different and non-inclusion variable sets. The other, non-chained variables are called \emph{isolated variables}. Let us regard a query clauses as a hypergraph where i) literals are hyperedges and ii) variables and constants are vertices. Fig. \ref{fig:queries} depicts query clauses $Q_a = \lnot A_1(x_1,x_2) \lor \lnot B(x_2,x_3) \lor \lnot C(x_3, x_4, x_5) \lor \lnot D(x_5, x_6) \lor \lnot E(x_3, x_4)$ and $Q_c = \lnot A_1(x_1,x_2,x_3) \lor \lnot B(x_3,x_4,x_5) \lor \lnot C(x_5, x_6, x_7) \lor \lnot D(x_1, x_7,x_8) \lor \lnot E(x_3, x_4, x_9)$ as hypergraphs and identifies chained variables and isolated variables in these query clauses. 


\textbf{Sep} simplifies a query clause by `cutting branches' from it. Without altering the computation of \textbf{Sep} in Section \ref{sec:tinf}, we specify the separation conditions in a goal-oriented way. A query clause is separated using
\begin{displaymath}
 \prftree[r,l]{\quad 
 \mbox{\vbox{\noindent
 if i) $A$ contains both isolated variables and chained variables, \\ ii) $\overline x = \Var(A) \cap \Var(D)$, iii) $\Var(C) \subseteq \Var(A)$, iv) $d_s$ is a fresh \\ predicate symbol, called as a definer.}}}{\textbf{Sep}: \ }
  {N \cup \{C \lor A \lor D\}}
  {N \cup \{C \lor A \lor d_s(\overline x), \lnot d_s(\overline x) \lor D\}}
\end{displaymath}
If a query clause $Q$ contains a literal $L$ where both isolated variables and chained variables occur, then \textbf{Sep} is iteratively applied to remove $L$ from $Q$, and terminates whenever $Q$ contains no such literal. Since query clauses $Q$ contain only negative literals, \textbf{Sep} separates $Q$ only into Horn clauses, i.e., clauses containing at most one positive literal. We sometimes omit `Horn' in following discussion since it is not critical. If \textbf{Sep} is applicable to a query clause, then it derives a query clause and a guarded clause (\textsc{lemma \ref{lem:sep_conlusion}}).

\begin{figure}[h]
\captionsetup{singlelinecheck=off}
\begin{minipage}[c]{0.46\textwidth}
  \includegraphics[width=\textwidth]{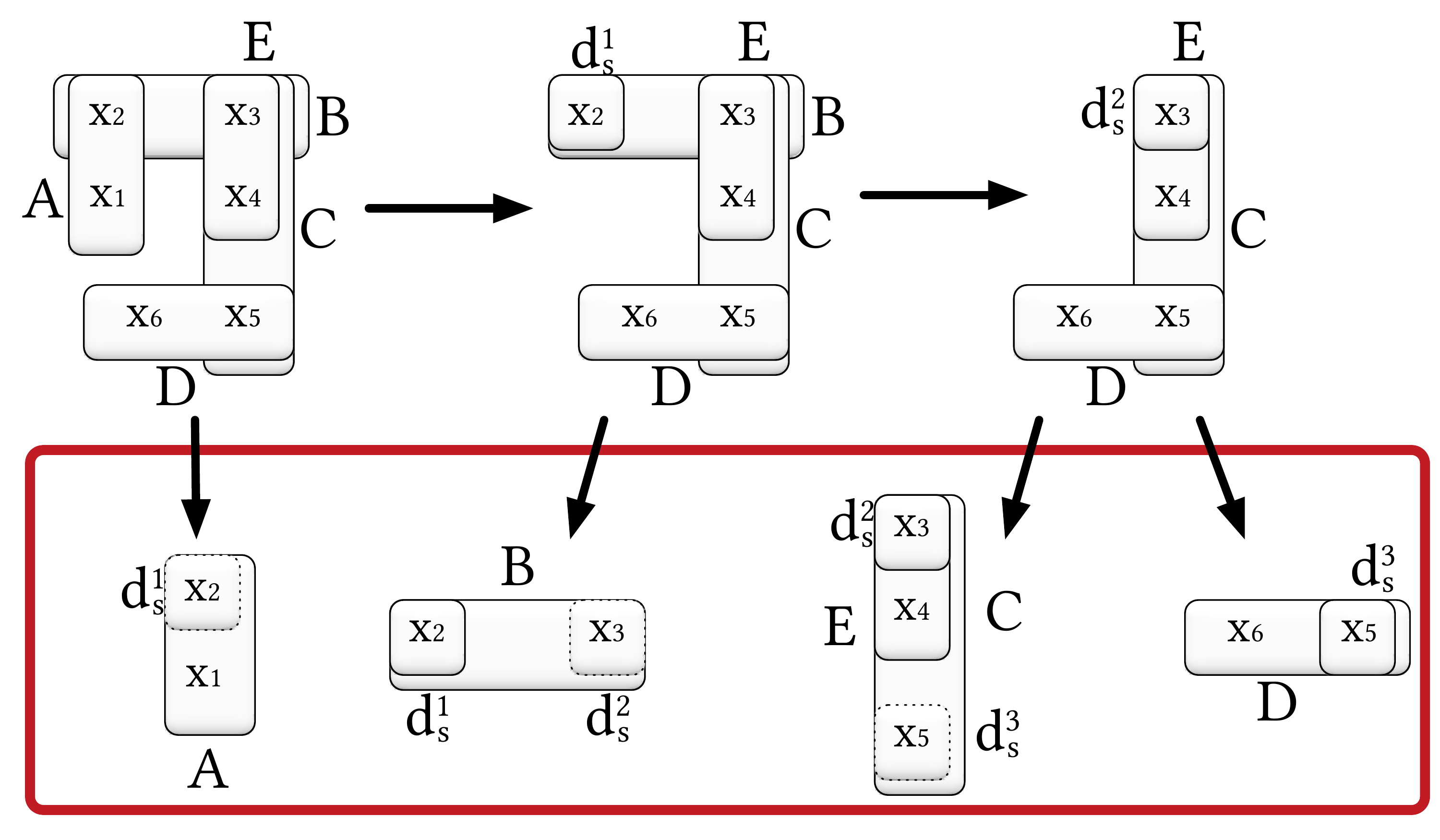}
 \end{minipage}\hfill
 \begin{minipage}[c]{0.5\textwidth}
	\caption[c]{Recursive applications of \textbf{Sep} to $Q_a$ transforms $Q_a$ into Horn guarded clauses 
	\begin{itemize}
	\item $\lnot A(x_1, x_2) \lor d_{s}^1(x_2)$,
	\item $\lnot B(x_2, x_3) \lor \lnot d_{s}^1(x_2) \lor d_{s}^2(x_3)$,
	\item $\lnot D(x_5, x_6) \lor \lnot d_{s}^3(x_5)$,
	\item $\lnot C(x_3,x_4,x_5) \lor \lnot E(x_3,x_4) \lor \lnot d_{s}^2(x_3) \lor d_{s}^3(x_5)$.
	\end{itemize}
	}
	\label{fig:sep_acyclic}
 \end{minipage}
\end{figure}

\begin{figure}[h]
\captionsetup{singlelinecheck=off}
\begin{minipage}[c]{0.46\textwidth}
  \includegraphics[width=\textwidth]{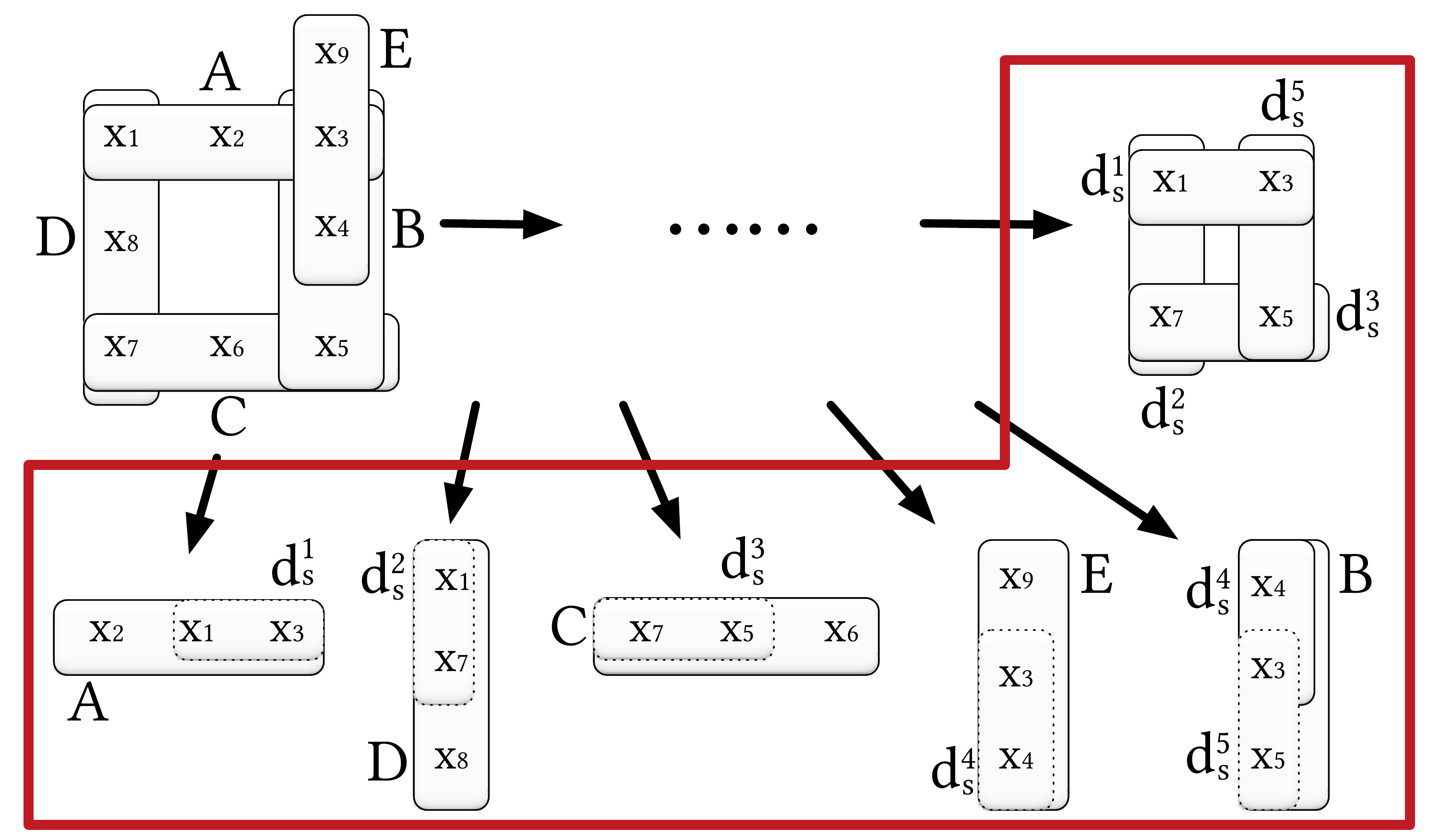}
 \end{minipage}\hfill
 \begin{minipage}[c]{0.5\textwidth}
	\caption[l]{Recursive applications of \textbf{Sep} to $Q_c$ transforms $Q_c$ into Horn guarded clauses
	\begin{itemize}
	\item $\lnot A(x_1, x_2, x_3) \lor d_{s}^1(x_1, x_3)$,
	\item $\lnot D(x_1, x_7, x_8) \lor d_{s}^2(x_1, x_7)$,
	\item $\lnot C(x_5 ,x_6 ,x_7) \lor d_{s}^3(x_5, x_7)$,
	\item $\lnot E(x_3 ,x_4 ,x_9) \lor d_{s}^4(x_3, x_4)$,
	\item $\lnot B(x_3, x_4, x_5) \lor \lnot d_{s}^4(x_3, x_4) \lor d_{s}^5(x_3, x_5)$,
	\end{itemize}
	and a chain-only query clause \\
	$\lnot d_{s}^1(x_1, x_3) \lor \lnot d_{s}^5(x_3, x_5) \lor \lnot d_{s}^3(x_5, x_7) \lor \lnot d_{s}^2(x_1, x_7)$.}
	\label{fig:sep_cyclic}
 \end{minipage}
\end{figure}

\textbf{Sep} is not applicable a query clause if it contains only chained variables or only isolated variables. We say query clauses containing only chained variables are \emph{chained-only query clauses} and containing only isolated variables are \emph{isolated-only query clauses}. E.g., $\lnot A(x_1,x_2) \lor \lnot B(x_2,x_3) \lor \lnot C(x_3,x_1)$ is a chained-only query clause where $\{x_1, x_2, x_3\}$ are chained variables, whereas $\lnot A(x_1,x_2,x_3) \lor \lnot B(x_2,x_3)$ is an isolated-only query clause where $\{x_1, x_2, x_3\}$ are isolated variables. Then by exhaustively applying \textbf{Sep} to a query clause, one obtains i) a set of guarded clauses and ii) a chained-only query clause or an isolated-only query clause. An indecomposable query clause is an isolated-only query clause iff it is a guarded clause (\textsc{lemma \ref{lem:isolated_only}}). Hence \textbf{Sep} transforms a query clause $Q$ into a set of guarded clauses and possibly a chained-only query clauses. Fig.~\ref{fig:sep_acyclic} and Fig. \ref{fig:sep_cyclic} show the iterative applications of \textbf{Sep} to query clauses $Q_a$ and $Q_c$ from Fig. \ref{fig:queries}, respectively. Fig. \ref{fig:sep_acyclic} shows that \textbf{Sep} transforms $Q_a$ into a set of Horn guarded clauses, while Fig. \ref{fig:sep_cyclic} shows that \textbf{Sep} transforms $Q_c$ into a set of Horn guarded clauses and a chained-only query clause. The dotted hypergraphs indicate positive literals and the undotted hypergraphs indicate negative ones. The hypergraphs of the resulting clauses are in the red boxes.

Interestingly, \textbf{Sep} (and \textbf{Split}) handles query clauses similarly as the so-called GYO-reduction in \cite{yu1979tree}. GYO-reduction recursively removes `ears' in the hypergraph of queries, to identified cyclic queries \cite{yannakakis1981acyclic}. Using GYO-reduction, a query $q$ is reduced to an empty query if $q$ is acyclic. In our context of query clauses, `ears' map to literals containing both isolated variables and chained variables, therefore are removed by \textbf{Sep} (and \textbf{Split}) from query clauses. Hence one can regard \textbf{Sep} (and \textbf{Split}) as a syntactical implementation of GYO-reduction, so that one can use \textbf{Sep} (and \textbf{Split}) to check cyclicity in queries (\textsc{lemma \ref{lem:GYO}}). 
%

\textbf{Sep} removes at most $n$ literals from an $n$-length query clause $Q$, and each application of \textbf{Sep} removes at least one literal from $Q$ without introducing new literals that contain both chained variables and isolated variables in the conclusion, hence, \textbf{Sep} can be applied at most linearly often to any query clause. 


\subsection*{\textbf{TRes} and \textbf{T-Trans} to handle chained-only query clauses}
In Fig. \ref{fig:sep_cyclic}, the chained-only query clause $\lnot d_{s}^1(x_1, x_3) \lor \lnot d_{s}^5(x_3, x_5) \lor \lnot d_{s}^3(x_5, x_7) \lor \lnot d_{s}^2(x_1, x_7)$  contains a cycle over $\{x_1, x_3, x_5, x_7\}$. We use the resolution rule \textbf{TRes}, under the refinement \emph{T-Refine}, to break such `variable cycles'. \textsc{Example \ref{example:chain-only}} shows how \textbf{TRes} breaks the cycle of $\{x_1, x_3, x_5, x_7\}$ while avoiding term depth growth in the resolvent.


\begin{example}
\label{example:chain-only}
Consider a chained-only query clause $Q$ and a set of guarded clauses $C_1, \ldots, C_4$:
\begin{gather*}
Q = \lnot d_{s}^1(x_1, x_3) \lor \lnot d_{s}^5(x_3, x_5) \lor \lnot d_{s}^3(x_5, x_7) \lor \lnot d_{s}^2(x_1, x_7) \qquad C_1 = d_{s}^1(x, gxy)^\ast \lor \lnot G_1xy \\
C_2 = d_{s}^5(gxy, x)^\ast \lor P(hxy) \lor \lnot G_2xy \qquad C_3 = d_{s}^3(fx, x)^\ast \lor \lnot G_3x \qquad C_4 = d_{s}^2(fx, x)^\ast \lor \lnot G_4x
\end{gather*}
\textbf{TRes} under the refinement \emph{T-Refine} is applicable to $Q, C_1, \ldots, C_4$. Then $\ComputeTop(Q, C_1, \ldots, C_4)$ computes the mgu $\sigma^\prime = \{x_1/fx, \ x_3/g(fx,y), \ x_5/fx, \ x_7/x\}$ to find top variables in $Q$. $x_3$ is the only top variable, so that $\SelectT$ selects $\lnot d_{s}^1(x_1, x_3)$ and $\lnot d_{s}^5(x_3, x_5)$, and \textbf{TRes} is performed on $Q$, $C_1$ and $C_3$, deriving $R = \lnot G_1xy \lor \lnot G_2xy \lor P(hxy) \lor \lnot d_s^3(x, x_7) \lor \lnot d_s^2(x_7, x)$. Notice that $R$ does not contain variable cycles since $\lnot G_1xy \lor \lnot G_2xy \lor P(hxy)$ in $R$, which replace $\lnot d_{s}^1(x_1, x_3) \lor \lnot d_{s}^5(x_3, x_5)$ in $Q$, is a guarded clause that does not maintain or expand the variable cycle $\{x_1, x_3, x_5, x_7\}$ in $Q$.
\end{example}

\begin{figure}[h]
\captionsetup{singlelinecheck=off}
\begin{minipage}[c]{0.46\textwidth}
  \includegraphics[width=\textwidth]{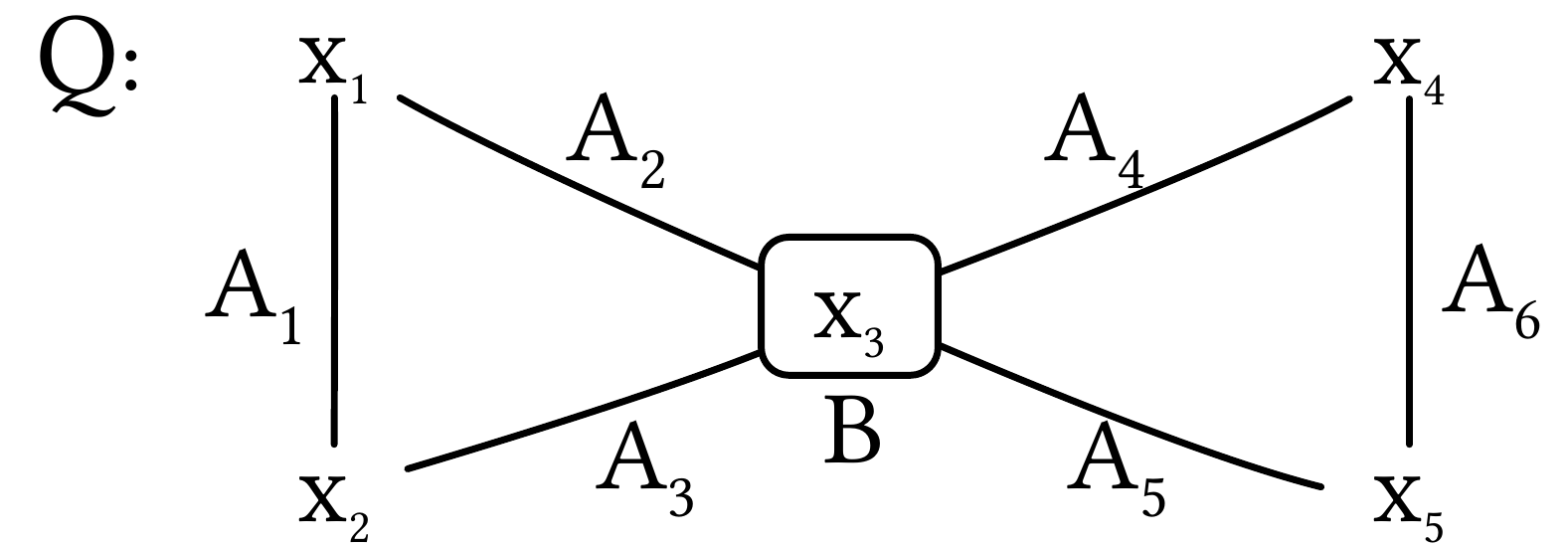}
 \end{minipage}\hfill
 \begin{minipage}[c]{0.5\textwidth}
	\caption[l]{Hypergraphs of $Q = \lnot A_1(x_1, x_2) \lor \lnot A_2(x_1, x_3) \lor \lnot A_3(x_2, x_3) \lor \lnot A_4(x_3, x_4) \lor \lnot A_5(x_3, x_5) \lor \lnot A_6(x_4, x_5) \lor \lnot B(x_3)$. $Q$ contains two variable cycles: $\{x_1, x_2, x_3\}$ and $\{x_3, x_4, x_5\}$.
	}
	\label{fig:two_cycles}
 \end{minipage}
\end{figure}

In \textsc{Example \ref{example:chain-only}}, \textbf{TRes} computes the top variables, in the positions in the main premise where the unification (Step 1 in $\ComputeTop$) `peaks'. The main premise in \textsc{Example \ref{example:chain-only}} contains only one `peak', therefore only single (loosely) guarded clauses occur in the resolvent. However, there can be multiple `peaks' that occur in different variable cycles, e.g., Fig. \ref{fig:two_cycles} illustrates a chained-only query clause $Q$ containing two different variable cycles. In a possible application of \textbf{TRes} to $Q$, if $\{x_1, x_2, x_4, x_5\}$ are only top variables (`peaks'), then two different (loosely) guarded clauses are introduced in the \textbf{TRes} resolvent, breaking cycles of $\{x_1, x_2, x_3\}$ and $\{x_3, x_4, x_5\}$, respectively (see details in \textsc{Example} \ref{example:multicycles} in \textbf{Appendix \ref{lem:handle_q}}).

The discussion above shows the case when several \emph{non-connected top variables} occur in different variable cycles. We say top variable $x_a$ and $x_b$ are \emph{connected} in a query clause $Q$ (w.r.t. an application of \textbf{TRes}) if there is a sequence of top variables $x_a, \ldots, x_b$ in $Q$ such that each adjacent pair of $x_i$ and $x_{i+1}$ in $x_a, \ldots, x_b$ occur in a literal of $Q$. Given a top variable $x$, \textbf{Algorithm \ref{algorithm:closed_top}} recursively finds the other top variables connected to $x$ by going through literals that contain $x$ and $x$-connected variables. In \textbf{Algorithm \ref{algorithm:closed_top}}, given a query clause $Q$, a top variable set $\mathcal{X}$ in $Q$ (w.r.t. an application of \textbf{TRes}) and a variable $x$ in $\mathcal{X}$, Lines~3--5 find the $x$-containing literals $\mathcal{L}$ in $Q$, and then identify top variables $\mathcal{X}^\prime$ in $\mathcal{L}$. The variables in $\mathcal{X}^\prime$ are connected to $x$ (and include $x$). Lines 6--7 check whether $\mathcal{X}^\prime$ are known $x$-connected variables: if so, all $x$-connected variables are found and the procedure terminates. In that case, $\mathcal{X}^\prime$ contains newly $x$-connected variables~$S_{temp}$, then \textbf{Algorithm \ref{algorithm:closed_top}} is run again to find top variables connected to variables in $S_{temp}$, and add them to the $x$-connected top variable set $S_{x}$.


\begin{algorithm}[h]
\DontPrintSemicolon
 \KwIn{A query clause $Q$, all top variables $\mathcal{X}$ in $Q$, and $x \in \mathcal{X}$}
%
%
%
%

 $S_{x} \gets \emptyset$, $S_{temp} \gets x$
 
 \While{$S_{temp} \not = \emptyset$}{
 
 Add all variables in $S_{temp}$ to $S_{x}$
 
 Find literals $\mathcal{L}$ in $Q$ where at least one variable in $S_{temp}$ occur
 
 Find all top variables $\mathcal{X}^\prime$ in $\mathcal{L}$
 
 \lIf{$\mathcal{X}^\prime \subseteq S_{x}$}
 {
 $S_{temp} \gets \emptyset$
 \tcp*[f]{No newly $x$-connected top variable, terminate}
 }
 \lElse{$S_{temp} \gets \mathcal{X}^\prime \setminus S_{x}$
 \tcp*[f]{Add newly $x$-connected top variables to $S_{x}$}}
 }
 {\Return $S_{x}$}
 
%
%
%
%
%
 
\caption{Compute an $x$-connected top variable set (a \emph{closed top variable set})}
\label{algorithm:closed_top}
\end{algorithm}

We call the output of \textbf{Algorithm \ref{algorithm:closed_top}} a \emph{closed top variables set} $\mathcal{X}$ (w.r.t. top variables in a chained-only query clause in an application of \textbf{TRes}), so that the variables in each pair of top variables in $\mathcal{X}$ are connected, and connected only by top variables in $\mathcal{X}$. In an application of \textbf{TRes} to a chained-only query clause $Q$ and (loosely) guarded clauses, if $\mathcal{X}$ is a closed top variable set, then all literals $\mathcal{L}$ (in $Q$) containing at least one variable in $\mathcal{X}$ will be replaced by a single (loosely) guarded clause in the \textbf{TRes} resolvent, and the remaining non-top-variable subclause in the \textbf{TRes} resolvent is a query clause (\textsc{lemma \ref{lem:tres_on_chain}}). To identify (loosely) guarded clauses in the \textbf{TRes} resolvent, we partition top variables $\mathcal{X}$ into disjoint closed top variable sets using \textbf{Algorithm \ref{algorithm:closed_top}}:
\begin{enumerate}
\item Pick a variable $x$ in $\mathcal{X}$, and apply \textbf{Algorithm \ref{algorithm:closed_top}} to $\mathcal{X}$ and $x$, obtaining a closed top variable set $\mathcal{X}_i$.
\item Let $\mathcal{X}$ be a new set that obtained by $\mathcal{X} \setminus \mathcal{X}_i$, and repeat Step 1 until $\mathcal{X}$ is empty.
\end{enumerate}
Then each pair of closed top variable sets have no overlapping top variable.

Resolvents obtained by applying \textbf{TRes} to a chain-only query clause and (loosely) guarded clauses is a disjunction of (loosely) guarded clauses and a query clause. Using the \textbf{T-Trans} technique, the \textbf{TRes} resolvent is transformed into a set of (loosely) guarded clauses and a query clause. For each closed top variable set $\mathcal{X}$ (in an application of \textbf{TRes} using (loosely) guarded clauses $\mathcal{S}$), \textbf{T-Trans} transforms the remainder of the $\mathcal{X}$-mapping side premises in $\mathcal{S}$ into a single (loosely) guarded clause (see \textsc{Example \ref{example:multicycles}} in \textbf{Appendix \ref{lem:handle_q}}).

\begin{definition}[\textbf{T-Trans}]
\label{def:ttrans}
Let \textbf{TRes} derive the resolvent $(\lnot A_{m+1} \lor \ldots \lor \lnot A_n \lor D_1 \lor \ldots \lor D_m)\sigma$ using (loosely) guarded clauses $B_1 \lor D_1, \ldots, B_n \lor D_n$, a chained-only query clause $Q = \lnot A_1 \lor \ldots \lor \lnot A_n$ and an mgu $\sigma$ such that $B_i\sigma = A_i\sigma$ for all $i$ such that $1 \leq i \leq m$. Then \textbf{T-Trans} introduces fresh predicate symbols $d_{t}$ in this manner: Let $\mathcal{X}_1, \ldots, \mathcal{X}_t$ be the partitioned top variable sets in $Q$. Then for each top variable set $\mathcal{X}_i$, we introduce a definer $\lnot d_{t}^i(\mathcal{X}_i)$ for $(D_{i_1} \lor \ldots \lor D_{i_k})\sigma$ ($\{i_1, \ldots, i_k\}$ is a subset of $\{1, \ldots, m\}$) if each literal in $\lnot A_{i_1} \lor \ldots \lor \lnot A_{i_k}$ contains at least one variable in $\mathcal{X}_i$.
\end{definition}

\textbf{T-Trans} is a renaming technique. Renaming is often used as a pre-processing approach to transform formulas into suitable clauses. When such renaming is used during the derivation, one needs to ensure only finitely many definers are introduced. \textbf{T-Trans} ensures this (\textsc{lemma \ref{lem:ttrans_conclusion}}). Note that in applications of \textbf{Sep} or \textbf{T-Trans}, we restrict that the introduced definer symbols in the conclusion are always smaller than predicate symbols in its the premises.

\section{Querying GF and LGF}
\label{sec:query_lgf}
Combining the results from Sections \ref{sec:decide_lgc}--\ref{sec:handle_query}, now we can discuss the query answering and rewriting procedure \textbf{Q-AR} for the (loosely) guarded fragment, which is described in \textbf{Algorithm~\ref{algorithm:query_AR}}. As discussed in Section \ref{sec:handle_query}, \textbf{Conden} and \textbf{Split} are used whenever possible. Let $q$ be a BCQ,~$\mathcal{D}$ be ground atoms and $\Sigma$ be a (loosely) guarded theory. Then \textbf{Q-AR} uses \textbf{Q-Trans} to transform $\Sigma$ and $q$ into (loosely) guarded clauses $\mathcal{S}$ and query clauses $Q$, respectively (Lines 1--2), reducing $\Sigma \cup \mathcal{D} \models q$ to checking $\mathcal{S} \cup \mathcal{D} \cup Q$. The answer for $q$ is `Yes' if $\mathcal{S} \cup \mathcal{D} \cup Q$ derives an empty clause, or `No' otherwise. The following discusses \textbf{Q-AR} by lines in \textbf{Algorithm \ref{algorithm:query_AR}}. Note that \textbf{Q-AR} immediately terminates and returns an `Yes' whenever an empty clause is derived. Line 4: we exhaustively apply \textbf{Sep} to $Q$, to transform $Q$ into Horn guarded clauses $\mathcal{S}_Q$ and a query clause $Q_{new}$. $Q_{new}$ is either an isolated-only query clause or a chained-only query (\textsc{Lemma \ref{lem:sep_conlusion}}). Lines 6--7: $Q_{new}$ is an isolated-only query clause implies that $Q$ is a Horn guarded clause (\textsc{Lemma \ref{lem:isolated_only}}), therefore we use the inference system \emph{T-Inf} to saturate (loosely) guarded clauses $\mathcal{S}_Q \cup \mathcal{S} \cup Q_{new}$, denoting the set $\Saturate(\mathcal{S}_Q \cup \mathcal{S} \cup Q_{new})$. \textbf{Q-AR} terminates with either an empty clause or a saturation of $\mathcal{S}_Q \cup \mathcal{S} \cup Q_{new}$. Lines 8--9: Suppose $Q_{new}$ is a chain-only query clause. Then the known (loosely) guarded clauses $\mathcal{S}_Q \cup \mathcal{S}$ are saturated first, and the saturation is denoted as $S_{sat}$. Lines~10--14: Suppose \textbf{TRes} side premises for resolving with $Q_{new}$ exist in $S_{sat}$, i.e., \textbf{TRes} is applicable to $Q_{new}$. We compute the \textbf{TRes} resolvent $R$ and transform $R$ into (loosely) guarded clauses $\mathcal{S}_R$ and a smaller query clauses $Q_R$ (\textsc{lemma \ref{lem:ttrans_conclusion}}). Then \textbf{Algorithm \ref{algorithm:query_AR}} process $Q_R$ again, where $Q_R$ is regarded as an input query and (loosely) guarded clauses $\mathcal{S}_R \cup \mathcal{S}_{sat}$ are regarded as input (loosely) guarded clauses. Line 15: Suppose \textbf{TRes} side premises for resolving with $Q_{new}$ do not exist in $\mathcal{S}_{sat}$. This means no inference can be performed on $Q_{new}$. Then \textbf{Q-AR} terminates with the clausal set $\mathcal{S}_{sat} \cup Q_{new}$.

\begin{algorithm}[h]
\DontPrintSemicolon
 \KwIn{a BCQ $q$, a set of (loosely) guarded formulas $\Sigma$ (and ground atoms $\mathcal{D}$)}
Apply \textbf{Q-Trans} to $q$, obtaining a query clause $Q$

Apply \textbf{Q-Trans} to $\Sigma$, obtaining (loosely) guarded clauses $\mathcal{S}$



 \While{$Q$ is not a guarded clause}{

  Recursively apply \textbf{Sep} to $Q$, obtaining Horn guarded clauses $\mathcal{S}_Q$ and a query clause $Q_{new}$
 
 \Switch{$Q_{new}$}{
 
 \lCase{$Q_{new}$ is an isolated-only query clause}{
  \tcp*[r]{$Q_{new}$ is guarded, by \textsc{Lemma \ref{lem:isolated_only}}}
 	{\Return $\Saturate(\mathcal{S}_Q \cup \mathcal{S} \cup Q_{new})$}       
     }
     
  \Case{$Q_{new}$ is a chained-only query clause}{
  	$\mathcal{S}_{sat} \gets \Saturate(\mathcal{S}_Q \cup \mathcal{S})$
 	
 	\If{\textbf{TRes} is applicable to $Q_{new}$ and clauses $C_1, \ldots, C_n$ in $\mathcal{S}_{sat}$}{
 	
  	$R \gets \TRes(C_1, \ldots, C_n, Q_{new})$
	
  	Apply \textbf{T-Trans} to $R$, obtaining (loosely) guarded clauses $\mathcal{S}_{R}$ and a query clause $Q_{R}$
  
  	$\mathcal{S} \gets \mathcal{S}_{R} \cup \mathcal{S}_{sat}$
  	\tcp*[r]{Combine obtained (loosely) guarded clauses}
  	$Q \gets Q_{R}$
  	\tcp*[r]{Apply \textbf{Sep}, \textbf{TRes} and \textbf{T-Trans} to reduce $Q_{R}$ again}
  }
  
  \lElse{\Return $\{\mathcal{S}_{sat} \cup Q_{new}\}$}
  }}}
\caption{Query answering and rewriting procedure \textbf{Q-AR}}
\label{algorithm:query_AR}
\end{algorithm} 

Let $\Sigma$ be a (loosely) guarded theory. Then \emph{T-Inf} (and \textbf{Q-Trans}) is sufficient to check satisfiability of~$\Sigma$. Hence we can assume the given (loosely) guarded theory $\Sigma$ is satisfiable. Now we discuss the case when \textbf{Q-AR} is a query answering algorithm and when it is a query rewriting algorithm. Ground atoms satisfy Condition 2 of (loosely) guarded formulas. Hence, if the input of \textbf{Q-AR} are ground atoms in $\mathcal{D}$, BCQs $q$ and a (loosely) guarded theory $\Sigma$, then \textbf{Q-AR} yields a BCQ answering algorithm. On the other hand, if the input are BCQs $q$ and a (loosely) guarded theory $\Sigma$, without $\mathcal{D}$, \textbf{Q-AR} yields a query rewriting algorithm. \textbf{Q-AR} transforms $\Sigma$ into (loosely) guarded clauses $\mathcal{S}$ and transforms $q$ into query clauses $Q$. In this query rewriting setting, \textbf{Q-AR} produces either i) an empty clause, meaning without considering datasets, the answer of $q$ is `Yes', or ii) a saturation $S_{sat}$ of $\mathcal{S} \cup Q$, meaning one needs to check satisfiability of $S_{sat}$ together with ground atoms $\mathcal{D}$. For checking $S_{sat} \cup \mathcal{D}$, we propose three possible solutions: 1) Immediately \textbf{Q-AR} solves it. 2) Alternatively, knowing $S_{sat}$ is satisfiable, we can use new selection functions on $S_{sat}$ to check $S_{sat} \cup \mathcal{D}$. For example, since $S_{sat}$ contains only (loosely) guarded clause, one can select the (loosely) guard in each clause in $S_{sat}$, so that no inference can be performed among clauses in $S_{sat}$, and inferences can only be performed between ground atoms and (loosely) guarded clauses in $S_{sat}$. Since (loose) guards contains all variables of that clause, resolving (loose) guards with ground atoms only produces ground clauses. This makes the reasoning task on $S_{sat} \cup \mathcal{D}$ easy. 3) Also, one can transform $S_{sat} \cup \mathcal{D}$ back to query evaluation problem $\mathcal{D} \models \Sigma_{q}$ where $\Sigma_{q}$ is obtained by unskolemising and negating clauses in $S_{sat}$. $S_{sat}$ contains only query clauses and (loosely) guarded clauses, therefore can be transformed into a union of first-order queries (\textsc{lemma \ref{lem:clause_formula}}). Then one can use model checking algorithms for $\mathcal{D} \models \Sigma_{q}$, i.e., checking whether a disjunction $\Sigma_{q}$ of first-order queries is entailed by ground atoms $\mathcal{D}$.



\textbf{Q-AR} is a goal-oriented procedure aiming to reduce the given query clause into smaller query clauses (\textsc{lemma \ref{lem:smaller_query}}) until deriving either an empty clause or a saturation. Knowing that \emph{T-Inf} guarantees termination in deciding (loosely) guarded clauses (\textsc{theorem \ref{thm:decide}}), we can show that \textbf{Q-AR} guarantees termination. There are only finitely many new query clause and (loosely) guarded clauses are derived when applying \textbf{Q-AR}, since only finitely many definers are introduced (\textsc{lemma \ref{lem:finite_definer}}).

\begin{theorem}
\textbf{Q-AR} is sound and complete inference system for BCQ answering and rewriting problems over GF and LGF.
\end{theorem}

\section{Discussions and open problems}
In this paper, we present the first, as far as we know, practical BCQ answering and rewriting procedure \textbf{Q-AR} for GF and LGF. During the investigation of querying for GF and LGF, we found that the same resolution techniques in automated reasoning are connected to techniques used in database research. Since the mainstream query answering procedure in the area of databases uses a tableau-like chase approach \cite{abiteboul1995chase}, it would be interesting to see how resolution techniques can equally transfer in the other direction. It would also be of interest to investigate using resolution to solve query-related problems such as query optimisation and query explanation. Another interesting question is: Acyclic queries are known tractable \cite{yannakakis1981acyclic}, and we show that \textbf{Sep} and \textbf{Split} transforms these queries into another tractable queries, namely bounded hypertree width queries (queries that are in the form of guarded clauses). Can we use these rules (or find new rules) to transform known tractable queries into other tractable queries such as the ones considered in \cite{max2013cq,GSS2001cq}. This would allow automated theorem provers to help to identify tractability in queries in practice.

There are still many open problems: i) can \textbf{Q-AR} query more expressive guarded logics, such as guarded negation fragment, and clique guarded negation fragment, where one needs to cope with equality and branched cliques? As yet there are no practical decision procedure and no query answering and rewriting procedure for these fragments. Adding new rules such as ordered paramodulation \cite{bachmair1990restrictions}, we are confident \textbf{Q-AR} can solve these problems. ii) Although using CQ to retrieve answers from GF is undecidable, can we use \textbf{Q-AR} to retrieve non-Boolean answers from fragments of GF, such as the expressive description logic $\mathcal{ALCHOI}$?

%
%
\clearpage

The following appendices provide detailed lemmas, proofs and examples for the paper. 
\appendix
\section{Section \ref{sec:pre}: Preliminaries}
\label{appen:pre}

A \emph{substitution} is a mapping defined on variables, where variables denoting terms are mapped to terms. By $E\sigma$ we denote the result of applying the substitution $\sigma$ to an expression $E$ and call $E\sigma$ an \emph{instance} of $E$. An expression $E^\prime$ is a \emph{variant} of an expression $E$ if there exists a variable substitution $\sigma$ such that $E^\prime\sigma = E\sigma$. A substitution $\sigma$ is a unifier of two terms $s$ and $t$ if $s\sigma = t\sigma$; it is a \emph{most general unifier} (\emph{mgu}), if for every unifier $\theta$ of $s$ and $t$, there is a substitution $\rho$ such that $\sigma \rho = \theta$. $\sigma \rho$ denotes the composition of $\sigma$ and $\rho$ as mappings. A \emph{simultaneous most general unifier} $\sigma$ is an mgu of two sequences of terms $s_1, \ldots, s_n$ and $t_1, \ldots, t_n$ such that $s_i\sigma = t_i\sigma$ for each $1 \leq i \leq n$. As common, we use the term mgu to denote the notion of simultaneous mgu. 

\section{Section \ref{sec:clause}: From logic fragments to clausal sets}
\label{appen:clause}

We assume that all free variables are existential quantified, and formulas are transformed into prenex normal form before Skolemisation \cite{nonnengart2001computing}. \textbf{Q-Trans} transforms of (loosely) guarded formulas proceeds using the following steps, which is illustrated on the guarded formula $F_6 = \exists x (Axy \land \forall z (Bxz \to \exists u Czu))$ from \textsc{Example \ref{example:GF}}.
\begin{enumerate}
\item Add existential quantifiers to all free variables in $F_6$. \qquad \qquad \qquad $\exists y \exists x (Axy \land \forall z (Bxz \to \exists u Czu))$
\item Rewrite $\to$ and $\leftrightarrow$ using disjunctions and negations, and transform $F_6$ into negation normal form, obtaining the formula $F_{nnf}$. \qquad \qquad \qquad \qquad \qquad \qquad \qquad \qquad \ $\exists y \exists x (Axy \land \forall z (\lnot Bxz \lor \exists u Czu))$
\item Introduce fresh predicate symbols $d_{\forall}^{i}$ for each universally quantified subformula, obtaining $F_{str}$. $\exists y \exists x (Axy \land d_{\forall}^{1}x), \quad \forall x (\lnot d_{\forall}^{1}x \lor \forall z (\lnot Bxz \lor \exists u Czu))$
\item Transform formulas in $F_{str}$ into prenex normal form before Skolemisation. Introduce Skolem constants $a$ and $b$ and a Skolem function $f$, obtaining $F_{sko}$. \qquad \qquad $Aab \land d_{\forall}^{1}a, \quad \forall xz (\lnot d_{\forall}^{1}x \lor \lnot Bxz \lor C(z,fxz))$
\item Drop all universal quantifiers and transform $F_{sko}$ into conjunctive normal form, obtaining the clauses: $Aab, \quad \ d_{\forall}^{1}a, \quad \ \lnot d_{\forall}^{1}x \lor \lnot Bxz \lor C(z,fxz)$.
\end{enumerate}

\begin{proposition}
\textbf{Q-Trans} transforms a (loosely) guarded formula into a set of (loosely) guarded clauses, and a BCQ into a query clause. 	
\end{proposition}
\begin{proof}
We explain how \textbf{Q-Trans} transforms a (loosely) guarded formula to a set of clauses satisfying conditions of (loosely) guarded clauses. The ground case is trivial. Let $C$ be a non-ground (loosely) guarded clause. Since $C$ is function-free, all compound terms in $C$ are Skolem function terms. Then $C$ is simple. Prenex normal form and then Skolemisation guarantee that every existential quantification $\exists y$ occurs in the form of $\forall \overline x \exists y \phi$. Then arguments of $\exists y$ Skolem function terms contains $\overline x$. Hence $C$ is covering. As for Condition 4 (and 5) in the definition of GF (and LGF), let a (loosely) guarded formula $F$ be of the form $\forall \overline x(G_1 \land ... \land G_n \to F_1)$ and $F_1$ be a (loosely) guarded formula. $F$ is loosely guarded when $n > 1$ and guarded when $n = 1$. Variables in $\lnot G_1, ..., \lnot G_n$ can either be bounded or free. According to Condition 5 of the definition of LGF, each bounded $x$ co-occurs with either a free or bounded variable $y$ in a $\lnot G_i$ in $\lnot G_1, ..., \lnot G_n$. Each pair of free variables in $\lnot G_1, ..., \lnot G_n$ co-occur in a \emph{definer} $d_j$. Therefore, each pair of variables in $C$ co-occur in at least one negative flat literal ($\lnot G_i$ and $\lnot d_j$) in $C$, and these negative flat literals are loose guards. If $n = 1$, each pair of variables in $C$ co-occur in at least one negative flat literal, hence guard. The clausal form of each BCQ clearly satisfies the definition of query clauses.
\end{proof}

\section{Section \ref{sec:tinf}: Top-variable Inference System}
\label{appen:tinf}

Let $\succ$ be a strict ordering, called a \emph{precedence}, on the symbols in \textbf{C}, \textbf{F} and \textbf{P}. An ordering~$\succ$ on expressions is \emph{liftable} if $E_1 \succ E_2$ implies $E_1\sigma \succ E_2\sigma$ for all expressions $E_1$, $E_2$ and all substitutions $\sigma$. An ordering $\succ$ on literals is \emph{admissible}, if i) it is well-founded and total on ground literals, and liftable, ii) $\lnot{A} \succ A$ for all ground atoms $A$, iii) if $B \succ A$, then $B \succ \lnot A$ for all ground atoms $A$ and $B$. A ground literal $L$ is (\emph{strictly}) $\succ$\emph{-maximal with respect to a ground clause}~$C$ if for any $L^\prime$ in $C$, $L \succeq L^\prime$ ($L \succ L^\prime$). A non-ground literal $L$ is (\emph{strictly}) $\succ$\emph{maximal with respect to a non-ground clause} $C$ if and only if there is a ground substitution $\sigma$ such that $L\sigma$ is (strictly) maximal with respect to $C\sigma$, that is, for all $L ^\prime$ in $C$, $L\sigma \succeq L^\prime\sigma$ ($L\sigma \succ L^\prime\sigma$). A \emph{selection function} $\Select(C)$ selects a possibly empty set of occurrences of negative literals in a clause $C$ with no other restriction imposed. Inferences are only performed on eligible literals. A literal $L$ is \emph{eligible} in a clause~$C$ if either nothing is selected by the selection function $\Select(C)$ and $L$ is a $\succ$-maximal literal with respect to $C$, or $L$ is selected by $\Select(C)$.

Let $N$ be a set of clauses. A ground clause $C$ is \emph{redundant with respect to} $N$ if there are ground instances $C_1\sigma, \ldots, C_n\sigma$ in $N$ such that $C_1\sigma, \ldots, C_n\sigma \models C$ and for each $C_i\sigma$ in $C_1\sigma, \ldots, C_n\sigma$, $C \succ C_i\sigma$. A non-ground clause $C$ is \emph{redundant with respect to} $N$ if every ground instance of $C$ is redundant with respect to $N$. An inference is redundant if one of the premises is redundant, or its conclusion is redundant or an element of $N$. We say $N$ is \emph{saturated up to redundancy} (with respect to ordered resolution and selection) if any inference from non-redundant premises in $N$ is redundant in $N$.

\textbf{Algorithm \ref{algorithm:refine}} ensures at least one top-variable literal is computed, formally stated as:
\begin{lemma}
\label{lem:algo2_eligible}
$\SelectT$ in \textbf{Algorithm \ref{algorithm:refine}} computes at least one eligible literal.
\end{lemma}
\begin{proof}
An input of $\SelectT$ is either flat loosely guarded clause or a query clause. Let $C$ be a flat loosely guarded clause or a query clause. If side premises of $C$ do not exist, all negative literals in $C$ are selected and therefore eligible (Lines 8 and 11 in \textbf{Algorithm \ref{algorithm:refine}}). If side premises $C_1, \ldots, C_n$ of $C$ exist, there is an mgu~$\sigma^\prime$ among $C_1, \ldots, C_n$ and $C$. There must exist at least one negative literal $L\sigma^\prime$ containing a deepest term in $C\sigma^\prime$. This implies that these literals are top-variable literals in $C$. $C$ contains at least one top-variable literal, thus at least one literal in $C$ is eligible.
\end{proof}


\begin{lemma}
\label{lem:algo1_eligible}
Let $C$ be either a query clause, or a guarded clause, or a loosely guarded clause. Then $C$ satisfies at least one condition in \textbf{Algorithm \ref{algorithm:refine}}, and \textbf{Algorithm \ref{algorithm:refine}} guarantees producing at least one eligible literal in $C$.
\end{lemma}
\begin{proof}
$C$ can be either i) ground, or ii) non-ground, simple but not flat, or iii) non-ground and flat. According to \textsc{Definitions \ref{def:query}--\ref{def:gc}}, $C$ cannot be both non-ground and positive. Hence in \textbf{Algorithm \ref{algorithm:refine}}, Line 1 covers Case i), Lines 2--3 cover Case ii), and Line 4--5 cover Case iii). In \textbf{Algorithm \ref{algorithm:refine}}, trivially, Lines 1--4 guarantee producing at least one eligible literal. For non-ground flat clauses satisfying Line 5, using \textsc{Lemma \ref{lem:algo1_eligible}}, \textbf{Algorithm \ref{algorithm:refine}} guarantees producing at least one eligible literal. 
\end{proof} 

The followings justify the application of a-posteriori checking.

\begin{lemma}
\label{lem:com_large}
In a covering clause $C$, let literal $L_1$ contain compound terms but literal $L_2$ does not contain compound terms. Then $L_1 \succ_{lpo} L_2$.
\end{lemma}
\begin{proof}
i) Let the compound term in $L_1$ be ground. Then $C$ is ground since $C$ is covering (\textsc{Remark} \ref{rem:covering_ground}). Hence $L_1 \succ_{lpo} L_2$ since $L_1$ contains non-constant function symbols but $L_2$ does not. ii) Let the compound term $t$ in $L_1$ be non-ground. Then $\Var(t) = \Var(L_1) = \Var(C)$ since $C$ is covering. $\Var(L_2) \subseteq \Var(L_1)$ and $L_1$ contains function symbols while $L_2$ does not. Thus $L_1 \succ_{lpo} L_2$ for each input clause.
\end{proof}

\textsc{Lemma \ref{lem:com_large}} implies that in a compound-term covering clause $C$, a maximal literal (w.r.t. $\succ_{lpo}$) is always one of compound-term literals, which implies that $\Max(C)$ in Line 3 in \textbf{Algorithm \ref{algorithm:refine}} returns one of compound-term literals in $C$. Using \textsc{Lemma \ref{lem:com_large}}, we can justify an a-priori checking in \emph{T-Inf}.

\begin{lemma}
\label{lem:apost}
Under the restrictions of \emph{T-Refine}, if an eligible literal $L$ is (strictly) maximal in a covering clause $C$, then $L\sigma$ is (strictly) maximal in $C\sigma$ for any substitution $\sigma$.
\end{lemma}
\begin{proof}
Maximality is decided for clauses in Lines 1 and 3 in \textbf{Algorithm \ref{algorithm:refine}}. i) Line 1 means $C$ is ground. Immediately the (strict) maximality of $L$ in $C$ implies (strict) maximality of $L\sigma$ in $C\sigma$ for any substitution $\sigma$. ii) Now assume $C$ satisfies the conditions in Line 3 in \textbf{Algorithm \ref{algorithm:refine}}, and $L^\prime$ is distinct from the maximal literal $L$. \textsc{Lemma \ref{lem:com_large}} shows that $C$ containing compound terms implies that $L$ contains compound terms~(indeed non-ground, \textsc{Remark \ref{rem:covering_ground}}), namely $t$. $C$ is covering, hence $\Var(t) = \Var(L) = \Var(C)$. Knowing $\Var(L^\prime) \subseteq \Var(L)$, if $L$ is (strictly) maximal in $C$, $L \succeq_{lpo} L^\prime$ ($L \succ_{lpo} L^\prime$) implies $L\sigma \succeq_{lpo} L^\prime\sigma$~($L\sigma \succ_{lpo} L^\prime\sigma$) for any substitution $\sigma$. Hence $L\sigma$ is (strictly) maximal in $C\sigma$.
\end{proof}

In the applications of inference rules in \cite{bachmair2001resolution}, (strictly) maximal literals are decided by \emph{a-posteriori checking} when one applies an mgu $\sigma$ to the premise $C$ first, and then regards the (strictly) maximal literals in $C\sigma$ as eligible literals. Although a-posteriori checking is essential to decide many first-order logic fragments \cite{fermuller2001decision}, it has the overhead on pre-computing unnecessary instantiations. Thanks to the covering property of our clausal class, it is possible to use \emph{a-priori checking} in \emph{T-Inf} (\textsc{Lemmas \ref{lem:com_large}--\ref{lem:apost}}). This means that the (strictly) maximal literals are decided before instantiations since \emph{T-Inf} guarantees producing the same conclusions as one would produce by using an a-posteriori checking.

The followings justify that any form of `partial inference' makes the `maximal selection resolution' related to it redundant.

\begin{proposition}[Bachmair \& Ganzinger \cite{bachmair2001resolution}]
\label{prop:bg_redun}
Let \textbf{SRes} denote the following standard ordered and selection-based resolution inference
\begin{align*}
 \prftree[l]{}
  {A_1 ^\ast \lor D_1, \ \ldots, \ A_n ^\ast \lor D_n}
  { }
  {\boxed{\lnot A_1 \lor \ldots \lor \lnot A_n} \lor D}
  {D_1 \lor \ldots \lor D_n \lor D}
\end{align*}
where $L^\ast$ denotes the maximal literals, and $\boxed{\lnot L}$ denotes the selected literals. Let $\{1, \ldots, n\}$ be partitioned into two subsets $\{i_1, \ldots, i_k\}$ and $\{j_1, \ldots, i_h\}$, and let $N$ be a set of clause. Then an application of \textbf{SRes} is redundant in $N$ whenever the `partial conclusion'
\begin{align*}
\lnot A_{j_1} \lor \ldots \lor \lnot A_{j_h} \lor D_{i_1} \lor \ldots \lor D_{i_k} \lor D
\end{align*}
is implied by $A_1 \lor D_1, \ldots, A_n \lor D_n$ and finitely many clauses $\Delta$ in $N$ that are all smaller than $\lnot A_1 \lor \ldots \lor \lnot A_n \lor D$.
\end{proposition}
\begin{proof}
W.l.o.g., let the main premise be of the form $\lnot A_1 \lor \ldots \lor \lnot A_m \lor \lnot A_{m+1} \lor \ldots \lor \lnot A_n \lor D$ and the `partial conclusion' be $\lnot A_{m+1} \lor \ldots \lor \lnot A_n \lor D_1 \lor \ldots \lor D_m \lor D$ where $m < n$.

We know
\begin{align}
& A_1 \lor D_1, \ldots, A_n \lor D_n, \Delta \models \\
& \lnot A_{m+1} \lor \ldots \lor \lnot A_n \lor D_1 \lor \ldots \lor D_m \lor D.
\end{align}
Suppose $I$ is an interpretation of (B.1) and suppose $I \models A_1, \ldots, A_n$. Then
\begin{align}
I \models A_1, \ldots, A_n, A_1 \lor D_1, \ldots, A_n \lor D_n, \Delta.
\end{align}
Since $I \models A_{m+1}, \ldots, A_n$, from $I \models (2)$, we have
\begin{align}
I \models D_1 \lor \ldots \lor D_m \lor D.
\end{align}
Then we get 
\begin{align}
& A_1, \ldots, A_n, A_1 \lor D_1, \ldots, A_n \lor D_n, \Delta \\
& \models D_1 \lor \ldots \lor D_m \lor D.
\end{align}
(B.6) is a subclause of $D_1 \lor \ldots \lor D_n \lor D$ since $m < n$, and hence
\begin{equation}
\begin{aligned}
& A_1, \ldots, A_n, A_1 \lor D_1, \ldots, A_n \lor D_n, \Delta \\
& \models D_1 \lor \ldots \lor D_n \lor D 
\end{aligned}
\end{equation}
\textbf{SRes} is redundant in $N$ if 
\begin{equation}
\begin{aligned}
& A_1 \lor D_1, \ldots, A_n \lor D_n, \Delta \models D_1 \lor \ldots \lor D_n \lor D 
\end{aligned}
\end{equation}
Let $I$ be an arbitrary model satisfying that
\begin{align}
& I \models A_1 \lor D_1, \ldots, A_n \lor D_n, \Delta, \\
& \text{but} \ I \not \models D_1 \lor \ldots \lor D_n \lor D.
\end{align}
(B.10) implies $I \not \models D_1, \ldots, I \not \models D_n$, therefore, considering (B.9) we get
\begin{align}
I \models A_1, \ldots, A_n, \Delta.
\end{align}
From (B.9) and (B.11), we obtain 
\begin{align}
I \models A_1, \ldots, A_n, A_1 \lor D_1, \ldots, A_n \lor D_n, \Delta.	
\end{align}
According to (B.7), (B.12) implies $I \models D_1 \lor \ldots \lor D_n \lor D$, which refutes (B.10).
%
%
%
%
\end{proof}

\begin{proposition}
\label{pro:pres}
Let \textbf{SRes} denote the following standard ordered and selection-based resolution inference
\begin{align*}
 \prftree[l]{}
  {A_1 ^\ast \lor D_1, \ \ldots, \ A_n ^\ast \lor D_n}
  { }
  {\boxed{\lnot A_1 \lor \ldots \lor \lnot A_n} \lor D}
  {D_1 \lor \ldots \lor D_n \lor D}
\end{align*}
where $L^\ast$ denotes the maximal literals, and $\boxed{\lnot L}$ denotes the selected literals. Let \textbf{PRes} denote the following `partial ordered resolution inference' of \textbf{SRes}
\begin{align*}
 \prftree[l]{}
  {A_i ^\ast \lor D_i, \ \ldots, \ A_j ^\ast \lor D_j}
  { }
  {\boxed{\lnot A_i \lor \ldots \lor \lnot A_j} \lor D ^\prime}
  {D_i \lor \ldots \lor D_j \lor D ^\prime}
\end{align*}
\begin{enumerate}
\setlength{\itemindent}{-1em}
\item where $A_i \lor D_i, \ldots, A_j \lor D_j$ is a subset of $A_1 \lor D_1, \ldots, A_n \lor D_n$, and
\item the main premise $\lnot A_1 \lor \ldots \lor \lnot A_n \lor D$ of \textbf{SRes} is the same as the main premise $\lnot A_i \lor \ldots \lor \lnot A_j \lor D ^\prime$ of \textbf{PRes}.
\end{enumerate}
Let $N$ be a set of clauses and $N_p$ be the set obtained by adding the resolvent $D_i \lor \ldots \lor D_j \lor D ^\prime$ of \textbf{PRes} to $N$. Then the particular application above of \textbf{SRes} is redundant in $N_p$.
\end{proposition}
\begin{proof}
Consider the application of \textbf{PRes} as specified in \textsc{Proposition \ref{pro:pres}}, and suppose $C_{p}$ is the resolvent $D_i \lor \ldots \lor D_j \lor D ^\prime$ obtained. Then $C_p$ is smaller than the main premise~$C$ of \textbf{SRes} (and \textbf{PRes}), since eligible literals $\lnot A_i$ in $C$ are replaced by respective $D_i$ satisfying that $A_i \succ D_i$. This means a `partial conclusion' $C_{p}$ is implied by $A_1 \lor D_1, \ldots, A_n \lor D_n$ and $C_{p}$ itself, which is smaller than $C$. By \textsc{Proposition \ref{prop:bg_redun}}, the specified applications of \textbf{SRes} is redundant in $N_{p}$.
\end{proof}

\textsc{Proposition \ref{pro:pres}} means that whenever one uses selection-based resolution \textbf{SRes} and selects multiple negative literals $\mathcal{L}$ in the main premise, one can select a subset $\mathcal{L^\prime}$ of $\mathcal{L}$, and resolve the main premise with~$\mathcal{L^\prime}$'s corresponding side premises in \textbf{SRes}, to make \textbf{SRes} redundant. This idea is also given in paragraphs appended \textsc{Proposition} 5.10 in \cite{bachmair2001resolution} or, more straightforwardly, \textsc{Proposition} 6.5 in \cite{bachmair1997theory}.

Since we employ admissible orderings and selection functions such that each non-ground computations of \textbf{SRes} and \textbf{PRes} can be mapped to their corresponding ground computations with the corresponding eligible literals, \textsc{Proposition \ref{pro:pres}} can be lifted to first-order logic using the Lifting Lemma in \cite{bachmair2001resolution}.

\begin{lemma}
\label{lem:sep_sound}
The \textbf{Sep} premise $C \lor D$ is satisfiable iff the \textbf{Sep} conclusions $\lnot d_s(\overline x) \lor C$ and $d_s(\overline x) \lor D$ are satisfiable.
\end{lemma}
\begin{proof}
Recall the separation rule:
\begin{displaymath}
 \prftree[r,l]{\qquad \ \ 
 \mbox{\vbox{\noindent
 if i) $C \lor D$ is separable into $C$ and $D$, ii) $\overline x = \Var(C) \cap \Var(D)$, \\ iii) $d_s$ is a fresh predicate symbol.}}}{\textbf{Sep}: \ }
  {N \cup \{C \lor D\}}
  {N \cup \{\lnot d_s(\overline x) \lor C, \ d_s(\overline x) \lor D\}}
\end{displaymath}

If: Apply unrefined resolution to $\lnot d_s(\overline x) \lor C$ and $ d_s(\overline x) \lor D$ with eligible literals $\lnot d_s(\overline x)$ and $d_s(\overline x)$ respectively, one derives $C \lor D$. This means if there is an interpretation $I$ such that $I \models N \cup \{\lnot d_s(\overline x) \lor C\}$ and $I \models N \cup \{d_s(\overline x) \lor D\}$, then $I \models N \cup \{C \lor D\}$.

Only if: We prove this direction by showing that if there is an interpretation of $C \lor D$ can be extended to~$I^\prime$ such that $I^\prime \models \lnot d_s(\overline x) \lor C$ and $I^\prime \models d_s(\overline x) \lor D$. Let $C \lor D$ be satisfiable, $I$ be an interpretation of $N$ and~$C \lor D$ and $\sigma$ an arbitrary ground substitution over the Herbrand universe such that $I \models C\sigma \lor D\sigma$. Let~$I^\prime$ be an extension of $I$ such that $I^\prime \models d_s(\overline x\sigma)$ if and only if $I \models C\sigma$. $I^\prime$ is a model of the \textbf{Sep} premise, and we show that $I^\prime$ is also a model of the \textbf{Sep} conclusion. First we show $I^\prime \models (\lnot d_s(\overline x) \lor C)\sigma$ by showing the opposite: assume there is a ground substitution $\theta$ satisfies $I^\prime \not \models (\lnot d_s(\overline x) \lor C)\theta$. Hence $I^\prime \models d_s(\overline x\theta)$ and $I^\prime \not \models C\theta$. This contradicts that $C\theta$ and $d_s(\overline x\theta)$ are equisatisfiable under $I^\prime$. Hence $I^\prime \models N \cup \{\lnot d_s(\overline x) \lor C\}$ Likewise, we show $I^\prime \models (d_s(\overline x) \lor D)\sigma$ by showing the opposite: assume there is a ground substitution $\eta$ satisfies $I^\prime \not \models (d_s(\overline x) \lor D)\eta$. Then $I^\prime \not \models d_s(\overline x\eta)$ and $I^\prime \not \models D\eta$. $I^\prime \not \models d_s(\overline x\eta)$ implies that $I^\prime \not \models C\eta$ since $C\eta$ and $d_s(\overline x\eta)$ are equisatisfiable under $I^\prime$. Then $I^\prime \not \models C\eta \lor D\eta$, which refutes $I \models N \cup \{C \lor D\}$. Hence $I^\prime \models N \cup \{d_s(\overline x) \lor D\}$. The proofs are similar if either $C$ or $D$ or both occurs negatively.
\end{proof}

\begin{theorem} [Soundness of \emph{T-Inf}]
\label{thm:tinf_sound}
\emph{T-Inf} is a sound inference system.
\end{theorem} 
\begin{proof}
\textsc{lemma \ref{lem:sep_sound}} shows that \textbf{Sep} is a sound rule. The conclusions of \textbf{Deduct}, \textbf{Fact}, \textbf{Res}, \textbf{TRes}, \textbf{Conden} and \textbf{Delete} are logical consequences of their premises, therefore these rules are sound.
\end{proof}

\begin{theorem}[Refutational Completeness of \emph{T-Inf}]
\label{thm:tinf_refu}
Let $N$ be a set of clauses saturated up to redundancy with respect to \emph{T-Inf}. Then provided finitely many definers are introduced by \textbf{Sep} and structural transformations, $N$ is unsatisfiable if and only if $N$ contains the empty clause.
\end{theorem}
\begin{proof}
If: $N$ contains an empty clause means that there is no Herbrand model satisfying $N$. Then $N$ is unsatisfiable.

Only if: Assume a saturated set $N$ is unsatisfiable, but contains no empty clause. Then $N$ contains at least one minimal counterexample, namely $C$. Now we show that using any conclusion-generating rule in \emph{T-Inf}, $C$ can always be reduced to a smaller counterexample. One can regard \textbf{TRes} as a redundant elimination rule such that whenever the `maximal selection resolution' is applicable, \textbf{TRes} is used instead. Then according to the framework in \cite{bachmair2001resolution}, applying (whenever applicable) \textbf{Split}, \textbf{Conden}, \textbf{Fact}, \textbf{Res} and `maximal selection resolution' to $C$ always produces a smaller counterexample, thus refutes the minimality of~$C$.

News rules comparing to the framework in \cite{bachmair2001resolution} is \textbf{Sep}. Let \textbf{Sep} applicable to $C$, producing $C_1$ and $C_2$. \textbf{Sep} is sound and $C$ is a minimal counterexample, then either $C_1$ and $C_2$ is a counterexample. Definers in $C_1$ and $C_2$ are smaller than predicate symbols in $C$, so that $C_1$ and $C_2$ are smaller than $C$. Hence, at least one of $C_1$ and $C_2$ refutes the minimality of $C$.
\end{proof}

According to \textsc{Definitions} \ref{def:query}--\ref{def:gc}, constants are freely allowed in query clauses and (loosely) guarded clauses. As for ground compound terms, the covering property implies that
\begin{remark}
\label{rem:covering_ground}
If $C$ is a covering clause containing a ground compound term, then $C$ is ground.
\end{remark}

\section{Section \ref{sec:decide_lgc}: \emph{T-Inf} decides guarded clauses and loosely guarded clauses}
\label{appen:decide_lgc}

Considering the behaviour of \emph{T-Inf} rules on (loosely) guarded clauses, only the rules \textbf{Split}, \textbf{Sep}, \textbf{Fact}, \textbf{Conden}, \textbf{Res} and \textbf{TRes} in \emph{T-Inf} can derive new conclusions. However, \textbf{Split} and \textbf{Sep} are not applicable to (loosely) guarded clauses since these clauses are not separable. This is because given a guarded clause $C$, a guard contains all variables of $C$, hence one cannot partition $C$. If $C$ is loosely guarded, loose guards $\mathcal{G}$ contains all variables of $C$, and $\mathcal{G}$ cannot be partitioned due to the variable co-occurrence property (Condition~2b in \textsc{Definition} \ref{def:gc}). Hence, only the rules \textbf{Fact}, \textbf{Conden}, \textbf{Res} and \textbf{TRes} derive new conclusions from given (loosely) guarded clauses.

\begin{remark}
\label{rem:rem_lit}
Let $C$ be a (loosely) guarded clause and $C^\prime$ be a non-empty clause obtained by removing some literals from $C$. Then $C^\prime$ is a (loosely) guarded clause if $C^\prime$ contains (loosely) guards. 
\end{remark}
\begin{proof}
It is immediate that $C^\prime$ satisfies conditions of definitions of (loosely) guarded clauses.	
\end{proof}


\begin{lemma}
\label{lem:eligible_covering}
Under restrictions of \emph{T-Refine}, let $\mathcal{L}$ be eligible literals in a (loosely) guarded clause $C$. Then $\Var(\mathcal{L}) = \Var(C)$.
\end{lemma}
\begin{proof}
Under restrictions of \textbf{Algorithm \ref{algorithm:refine}}: i) if $C$ is ground (Line 1), then this Lemma immediately holds. ii) Suppose $C$ has compound terms (Lines 2--3). Then according to \textbf{Algorithm \ref{algorithm:refine}}, $C$ contains only one eligible literal $L$. $L$ is a compound-term literal by \textsc{Lemma \ref{lem:eligible_covering}} (if $L$ is positive) and the definition of $\SelectNC$~(if $L$ is negative). Let $t$ be a compound term in $L$. Since $\Var(t) = \Var(C)$ ($C$ is covering) and $\Var(t) = \Var(L)$ ($t$ occurs in $L$), $\Var(L) = \Var(C)$.	iii) Suppose $C$ is a flat guarded clause (Line 4). Since one of the guards $L$ is eligible and selected by $\SelectG$, according to the definition of guards in Condition 2 in \textsc{Definition} \ref{def:gc} , $\Var(L) = \Var(C)$. iv) Suppose $C$ is a flat loosely guarded clause (Line 5), and $\SelectT$ selects top-variable literals in $C$. W.l.o.g. assume $x$ is a top variable. According to Condition 3 of \textsc{Definition} \ref{def:gc}, $x$ co-occurs with all other variables in $C$ in some loose guards, which are indeed eligible top-variable literals~$\mathcal{L}$ since each of these loose guards contains $x$. Then $\Var(\mathcal{L}) = \Var(C)$. 
\end{proof}

\textsc{Lemma \ref{lem:eligible_covering}} implies that we only need to investigate the unification of eligible literals to understand unification of all variables in the premises. Given two expressions $A(\ldots, t, \ldots)$ and $B(\ldots, u, \ldots)$, we say $t$ \emph{matches} $u$ if the argument position of $t$ in $A$ is the same as that of $u$ in $B$. Then we discuss the unification of two eligible literals by explaining the matching arguments between these literals. \textsc{Lemma \ref{lem:mat_comp_lit}} describes the matching between two compound-term eligible literals. 

\begin{lemma}
\label{lem:mat_comp_lit}
Let $A_1(\ldots, t, \ldots)$ and $A_2(\ldots, u, \ldots)$ be two non-ground, simple and covering literals containing compound terms $t$ and $u$, respectively. Then $t$ matches $u$ if there is an mgu $\sigma$ between $A_1$ and~$A_2$.	
\end{lemma}
\begin{proof}
In the following proofs, whenever terms $t$, $t^\prime$, $u$ and $u^\prime$ are assumed to be compound, immediately we conclude that they are non-ground (ground compound terms implies a covering literal/clause being ground, \textsc{Remark \ref{rem:covering_ground}}). Let $t$ be a compound term and $u$ be a non-compound term (hence a constant or a variable). i) $u$ being a constant immediately fails the unification $t\sigma = u\sigma$. ii) Let $u$ be a variable. Assume $u^\prime$ is the compound term in $A_2$ that matches non-ground term $t^\prime$ in $A_1$. The unification $t\sigma = u\sigma$ implies that the variable $u$ is unified as a compound term $t\sigma$. Then since $u \in \Var(u^\prime)$, $u^\prime\sigma$ is a nested compound term and $\Var(t) \subseteq \Var(u^\prime\sigma)$. If $t^\prime$ is a variable, then $t^\prime \in \Var(t)$ and $\Var(t) \subseteq \Var(u^\prime\sigma)$ means $t^\prime \in \Var(u^\prime\sigma)$, thus fails the unification of $t^\prime\sigma = u^\prime\sigma$. If $t^\prime$ is a compound term, then $\Var(t^\prime) = \Var(t)$ and $\Var(t) \subseteq \Var(u^\prime\sigma)$ means $\Var(t^\prime) \subseteq \Var(u^\prime\sigma)$, thus fails the unification of $t^\prime\sigma = u^\prime\sigma$. Hence, compound terms in $A_1$ and $A_2$ match each other.
\end{proof}

Eligible literals restrict instantiations of (loosely) guards as follows: 

\begin{lemma}
\label{lem:guard_tinf}
Let two simple and covering atoms (flat literals are implicitly covering) $A_1$ and $A_2$ be unifiable using an mgu $\sigma$, and a flat literal $G$ and flat literals $\mathcal{G}$ satisfy that $\Var(A_1) = \Var(G) = \Var(\mathcal{G})$. Then $G\sigma$ and $\mathcal{G}\sigma$ are flat and $\Var(A_1\sigma) = \Var(G\sigma) = \Var(\mathcal{G}\sigma)$ if $A_1$ is a compound-term atom.
\end{lemma}
\begin{proof}
$\Var(A_1) = \Var(G) = \Var(\mathcal{G})$ immediately implies that $\Var(A_1\sigma) = \Var(G\sigma) = \Var(\mathcal{G}\sigma)$. Now we show that $G\sigma$ and $\mathcal{G}\sigma$ are flat. $A_2$ can be either a compound-term or a flat atom. If $A_2$ is ground, since $A_1$ contains compound terms, then $A_2$ must contain ground compound terms that match compound terms in $A_1$, or else $A_1$ and $A_2$ are not unifiable. $A_1$ is covering and $A_1$ and $A_2$ are simple atoms, so $\sigma$ substitutes all variables in $A_1$ with constants. Now let $A_2$ be a compound-term atom. Then compound terms in $A_2$ are non-ground (\textsc{Remark} \ref{rem:covering_ground}). \textsc{Lemma \ref{lem:mat_comp_lit}} shows that compound terms in $A_1$ and $A_2$ match each other. Since~$A_1$ and $A_2$ are covering, and compound terms contains only variables and constant, $\sigma$ substitutes variables in $A_1$ with either variables or constants. Now let $A_2$ be a flat atom. Then $\sigma$ can only substitute variables in $A_1$ with either variables or constants. In all possible cases, $\sigma$ substitutes $A_1$ variables with either variables or constants. Then $\Var(A_1) = \Var(G) = \Var(\mathcal{G})$, and $G$ and $\mathcal{G}$ being flat imply that $G\sigma$ and $\mathcal{G}\sigma$ are flat.
\end{proof}

\textsc{Lemma 4.6} in \cite{ganzinger1999superposition} gives a similar result as \textsc{Lemma} \ref{lem:simple} below, but a key condition `covering' was not considered. Now using \textsc{Lemma \ref{lem:mat_comp_lit}}, we show that no term depth occurs after unification of eligible literals. 

\begin{lemma}
\label{lem:simple}
Let two simple and covering literals $A_1$ and $A_2$ be unifiable with an mgu $\sigma$. Then $A_1\sigma$ is simple.
\end{lemma}
\begin{proof}
If either of $A_1$ and $A_2$ is ground, or either of $A_1$ and $A_2$ is non-ground and flat, then immediately $A_1\sigma$ is simple. Let both $A_1$ and $A_2$ contain compound terms. W.l.o.g. let $t$ and $u$ be matching compound terms in $A_1$ and $A_2$, respectively (\textsc{Lemma \ref{lem:mat_comp_lit}}). Both $A_1$ and $A_2$ are simple implies that $t$ and $u$ contain either variables or constants. Then $\sigma$ substitutes variables with either constants or variables. Since $A_1$ is covering, $\Var(t) = \Var(A_1)$ implies that $\sigma$ substitutes all variables in $A_1$ with either constants or variables. Then $A_1$ is simple implies that $A_1\sigma$ is simple.
\end{proof}

\textsc{Lemmas \ref{lem:mat_comp_lit}--\ref{lem:simple}} consider the matchings and unifications between eligible literals. Eligible literals in guarded clauses and loosely guarded clauses limit instantiations of non-eligible literals, formally stated as:
\begin{lemma}
\label{lem:conclusion}
[Extends \textsc{lemma 4.7} in \cite{ganzinger1999superposition}]
Let $A_1$ and $A_2$ be two simple atoms and $\mathcal{A}$ be a set of simple atoms satisfying that 
\begin{itemize}
\item $\Var(A_2) \subseteq \Var(A_1)$ and $\Var(A_2) \subseteq \Var(\mathcal{A})$, and
\item if $A_1$, $A_2$ and $\mathcal{A}$ contain compound terms, namely $t$, $u$ and $s$ respectively, then $\Var(t) = \Var(u) = \Var(s)$.	
\end{itemize}
Then given an arbitrary substitution $\sigma$, these conditions hold:
\begin{enumerate}
\item If $A_1\sigma$ is simple or $\mathcal{A}\sigma$ is simple, $A_2\sigma$ is simple. 
\item $\Var(A_2\sigma) \subseteq \Var(A_1\sigma)$ and $\Var(A_2\sigma) \subseteq \Var(\mathcal{A}\sigma)$.
\item $\Var(t\sigma) = \Var(u\sigma) = \Var(s\sigma)$.
\end{enumerate}
\end{lemma}
\begin{proof}
Item 1: $A_1$ ($\mathcal{A}$) is simple. Hence $A_1\sigma$ ($\mathcal{A}\sigma$) is simple implies that $\sigma$ does not increase term depth in $A_1$ ($\mathcal{A}$). Since $\Var(A_2) \subseteq \Var(A_1)$ ($\Var(A_2) \subseteq \Var(\mathcal{A})$) and $A_2$ is simple, $A_2\sigma$ is simple. Item~2: $\Var(A_2) \subseteq \Var(A_1)$ ($\Var(A_2) \subseteq \Var(\mathcal{A})$) implies $\Var(A_2\sigma) \subseteq \Var(A_1\sigma)$ ($\Var(A_2\sigma) \subseteq \Var(\mathcal{A}\sigma)$). Item 3: $\Var(t) = \Var(u) = \Var(s)$ implies $\Var(t\sigma) = \Var(u\sigma) = \Var(s\sigma)$.
\end{proof}

\begin{remark}
Let $C$ be a (loosely) guarded clause and $C^\prime$ be a non-empty clause obtained by removing some literals from $C$. Then $C^\prime$ is a (loosely) guarded clause if $C^\prime$ contains (loosely) guards. 
\end{remark}
\begin{proof}
It is immediate that $C^\prime$ satisfies conditions of definitions of (loosely) guarded clauses.	
\end{proof}

\begin{lemma}
\label{lem:conden}
A condensation of a (loosely) guarded clause is a (loosely) guarded clause.
\end{lemma}
\begin{proof}
Let a (loosely) guarded clause $C$ be the premise in \textbf{Conden}, and $C^\prime$ be the condensation result of $C$. Since $C^\prime$ is a proper subclause of $C$ and \textbf{Conden} cannot remove all guards (all loose guards) in $C$, then according to \textsc{Remark} \ref{rem:rem_lit}, $C^\prime$ is a (loosely) guarded clause.
\end{proof}

\begin{lemma}
\label{lem:fact}
In the application of \textbf{Fact}, a factor of a (loosely) guarded clause is a (loosely) guarded clause.
\end{lemma}
\begin{proof}
Let a (loosely) guarded clause $C = A_1 \lor A_2 \lor D$ produces $C^\prime = (A_1 \lor D)\sigma$ using \textbf{Fact} where $A_1$ and $A_2$ are two eligible literals and $\sigma$ is an mgu between $A_1$ and $A_2$. Then $C^\prime$ is calculated when $C$ satisfies conditions of Line 1 or Line 3 in \textbf{Algorithm \ref{algorithm:refine}}. Line 1: $C$ is ground implies that $C^\prime$ is ground. Line 3: $C$ is not ground and contains positive compound-term literals. \textsc{Lemma \ref{lem:com_large}} implies that $A_1$ and $A_2$ are positive compound-term literals. Let $C$ contains a guard $G$ (loosely guards $\mathcal{G}$), a literal~$L$ and a compound term $t$. Since $C$ is a (loosely) guarded clause, $A_1$ and $A_2$ are simple and covering such that $\Var(A_1) = \Var(A_2) = \Var(C)$. According to \textsc{Lemma \ref{lem:simple}}, $A_1\sigma$ is simple. Then by Item 1 in \textsc{Lemma~\ref{lem:conclusion}}, since $L$ is simple and $\Var(L) \subseteq \Var(A_1)$, $L\sigma$ is simple. Hence $C\sigma$ is simple. $C$ is covering implies $\Var(A_1) = \Var(C) = \Var(t)$. Then Item 3 in \textsc{Lemma \ref{lem:conclusion}} shows that $\Var(C\sigma) = \Var(t\sigma)$. Hence $C\sigma$ is covering. Since $\Var(A_1) = \Var(G) = \Var(\mathcal{G}) = \Var(C)$ and $A_1$ is a compound-term literal, \textsc{Lemma \ref{lem:guard_tinf}} shows that $\Var(G\sigma) = \Var(C\sigma)$ ($\Var(\mathcal{G}\sigma) = \Var(C\sigma)$) and $G\sigma$ ($\mathcal{G}\sigma$) is flat, hence $C\sigma$ is (loosely) guarded by $G$ ($\mathcal{G}$). Then $C\sigma$ is (loosely) guarded. This implies that $C^\prime$ is a (loosely) guarded clause, shown by \textsc{Remark} \ref{rem:rem_lit}.
\end{proof}

\begin{lemma}
\label{lem:res}
In the application of \textbf{Res}, resolvents of (loosely) guarded clauses are (loosely) guarded clauses.
\end{lemma}
\begin{proof}
Let (loosely) guarded clauses $C_1 = A_1 \lor D_1$ and $C_2 = \lnot A_2 \lor D_2$ be the positive premise and the negative premise in \textbf{Res}, respectively, producing a resolvent $C^\prime = (D_1 \lor D_2)\sigma$, where $A_1$ and $\lnot A_2$ are eligible literals so that an mgu $\sigma$ satisfies $A_1\sigma = A_2\sigma$. Then according to \textbf{Algorithm \ref{algorithm:refine}}, $C_1$ satisfies a condition in one of Line 1 or 3, and $C_2$ satisfies a condition in one of Line 1, 2 or 4. 

Case 1: Let either $C_1$ or $C_2$ be ground (Line 1 in \textbf{Algorithm \ref{algorithm:refine}}). \textsc{Lemma \ref{lem:eligible_covering}} shows that if one of eligible literals in either $C_1$ or $C_2$ is ground, then $\sigma$ is a ground substitution so that $C_1\sigma$ and $C_2\sigma$ are ground. Then $C^\prime$ is ground, thus a (loosely) guarded clause. 

Case 2: Let $C_1$ contain compound terms (Line 3 in \textbf{Algorithm \ref{algorithm:refine}}) and $C_2$ either contain compound terms or be flat (Line 2 or 4 in \textbf{Algorithm \ref{algorithm:refine}}, respectively). Assume $G$ is a guard (or loose guards $\mathcal{G}$) in $C_1$, $L$ is a literal and $t$ is a compound term in $C_1$ or $C_2$. Since $A_1$ is eligible in a compound-term clause $C_1$, according to \textsc{Lemma \ref{lem:com_large}}, $A_1$ is a compound-term literal. Then $A_1$ and $A_2$ satisfies conditions in \textsc{Lemma \ref{lem:guard_tinf}}, so that $G\sigma$ and $\mathcal{G}\sigma$ are flat and $\Var(A_1\sigma) = \Var(G\sigma) = \Var(\mathcal{G}\sigma)$. \textsc{Lemma \ref{lem:eligible_covering}} shows that $\Var(A_1) = \Var(C_1)$, $\Var(A_2) = \Var(C_2)$, and $A_1$ and $A_2$ are unifiable. This implies that $\Var(A_1\sigma) = \Var(A_2\sigma)$, thus $\Var(A_1\sigma) = \Var(C_1\sigma) = \Var(C_2\sigma)$. Then $\Var(G\sigma) = \Var(\mathcal{G}\sigma) = \Var(C_1\sigma) = \Var(C_2\sigma)$. Hence $\Var(G\sigma) = \Var(\mathcal{G}\sigma) = \Var(C^\prime)$ and $G\sigma$ is a guard (or $\mathcal{G}\sigma$ are loose guards) in $C^\prime$. Since \textsc{Lemma~\ref{lem:eligible_covering}} shows $\Var(L) \subseteq \Var(A_1)$ (or $\Var(L) \subseteq \Var(A_2)$) and \textsc{Lemma} \ref{lem:simple} shows $A_1\sigma$ (or $A_2\sigma$) is simple, using Item 1 in \textsc{Lemma \ref{lem:conclusion}}, $L\sigma$ is simple. Hence $C^\prime$ is simple. Because that $C_1$ and $C_2$ are covering and \textsc{Lemma}~\ref{lem:eligible_covering}, we know that $\Var(t) = \Var(A_1)$ (or $\Var(t) = \Var(A_2)$). Then according to Item 5 in \textsc{Lemma \ref{lem:conclusion}}, $\Var(t\sigma) = \Var(A_1\sigma) = \Var(G\sigma) = \Var(\mathcal{G}\sigma)$. Hence $C^\prime$ is covering. Therefore $C^\prime$ is a (loosely) guarded clause.
\end{proof}

\textsc{Lemma \ref{lem:res}} shows \textbf{Res} computations on (loosely) guarded clauses. When a flat loosely guarded clause participates in a resolution computation, \textbf{TRes} is used to derive conclusions. First we discuss the unifications in \textbf{TRes}.
\begin{lemma}
\label{lem:tres_matching}
In an application of \textbf{TRes} to a flat clause as the main premise and a set of (loosely) guarded clauses as side premises, these conditions hold:
\begin{enumerate}
\item Top variables match constants or compound terms, and non-top-variables match constants or variables.
\item The \textbf{TRes} mgu $\sigma$ substitutes top variables $x$ with either constants or the compound term (modulo variable renaming) matching $x$, and substitutes non-top variables with either constants or variables.
\item Let a top variable $x$ match a constant. Then all negative literals in the main premise are selected and the \textbf{TRes} mgu $\sigma$ is a ground substitution that substitutes variables with only constants.
\end{enumerate}
\end{lemma}
\begin{proof}
Recall the definition of
\begin{align*}
 \prftree[l]{\textbf{TRes}: \quad}
  {B_1 \lor D_1 \ \ \ldots \ \ B_m \lor D_m \ \ \ldots \ \ B_n \lor D_n}
  { }
  { }
  { }
  {\lnot A_1 \lor \ldots \lor \lnot A_m \lor \ldots \lor \lnot A_n \lor D}
  {(D_1 \lor \ldots \lor D_m \lor \lnot A_{m+1} \lor \ldots \lor \lnot A_{n} \lor D)\sigma}
\end{align*}
where i) there exists an mgu $\sigma^\prime$ such that $B_i\sigma^\prime = A_i\sigma^\prime$ for each $i$ such that $1 \leq i \leq n$, making $\lnot A_1 \lor \ldots \lor \lnot A_m$ top-variable literals and being selected, and $D$ is positive, ii) no literal is selected in $D_1, \ldots, D_n$ and $B_1, \ldots, B_n$ are strictly $\succ$-maximal with respect to $D_1, \ldots, D_n$, respectively. $\sigma$ is an mgu such that $B_i\sigma = A_i\sigma$ for all such that $1 \leq i \leq m$.

Since each literal in $\lnot A_1 \lor \ldots \lor \lnot A_m$ contains at least one top variable, w.l.o.g. assume $\lnot A_t(\ldots, x, \ldots, y, \ldots)$ is a literal in $\lnot A_1 \lor \ldots \lor \lnot A_n$ such that $x$ is a top variable and $y$ is not a top variable (if such a variable exists). Suppose $C_t = B_t(\ldots, t_1, \ldots, t_2, \ldots) \lor D$ is a side premise such that $B_t(\ldots, t_1, \ldots, t_2, \ldots)$ matches $A_t(\ldots, x, \ldots, y, \ldots)$. We need to show that $t_1$ is either a constant or a compound term and $t_2$ is either a constant or a variable. 

Assume $C_t$ is ground. Then $t_1$ is a constant or a ground compound term. Assume $t_2$ is neither a constant nor a variable. Then $t_2$ is a ground compound term. Hence $\Dep(t_1) \geq \Dep(t_2)$. Since $t_1$ and $t_2$ are ground, $\Dep(t_1\sigma^\prime) \geq \Dep(t_2\sigma^\prime)$. Thus $\Dep(y\sigma^\prime) \geq \Dep(x\sigma^\prime)$. This contradicts $y$ not being a top variable. Now assume $C_t$ is not ground. Then according to \textbf{Algorithm \ref{algorithm:refine}}, $C_t$ contains compound-term literals (indeed non-ground, \textsc{Remark \ref{rem:covering_ground}}) or else at least one literal in $C_t$ would be selected. Since \textsc{Lemma \ref{lem:com_large}} implies that only the compound-term literal can be maximal, $B_t$ is a compound-term literal. Assume $t_1$ is not a compound term. Then there exists a compound term $t$ in $B_t$ that matches a variable $z$ in $A_t$. Since $\Var(t_1) \subseteq \Var(t)$~(by the covering property) and $\Dep(t_1) < \Dep(t)$, $\Dep(t_1\sigma^\prime) < \Dep(t\sigma^\prime)$. Hence, $\Dep(x\sigma^\prime) < \Dep(z\sigma^\prime)$. This contradicts $x$ being a top variable. Then $t_1$ must be a compound term. Now assume $t_2$ is neither a constant nor a variable. Then $t_2$ is a compound term. $\Var(t_1) = \Var(t_2)$ (the covering property) and $\Dep(t_1) = \Dep(t_2)$ imply $\Dep(t_1\sigma^\prime) = \Dep(t_2\sigma^\prime)$. Hence $\Dep(x\sigma^\prime) = \Dep(y\sigma^\prime)$. This contradicts $y$ not being a top variable. Hence, $y$ matches a constant or a variable. Such matchings immediately shows that $\sigma$ substitutes top variables with either constants or compound terms, and substitutes non-top variables with either constants or variables.

Suppose a top variable $x$ matches a constants. The definition of $\ComputeTop$ shows that for any non-top variable (if such a variable exists) $y$ in the main premise, $\Dep(x\sigma^\prime) > \Dep(y\sigma^\prime)$. $x$ matching constants indicates that $x\sigma^\prime$ is a constant, therefore $\Dep(x\sigma^\prime) = 0$. $\Dep(y\sigma^\prime)$ cannot be smaller than 0, hence, all variables in $\lnot A_1 \lor \ldots \lor \lnot A_n$ are top variables, and for any variable $x$ in $\lnot A_1 \lor \ldots \lor \lnot A_n$, $\Dep(x\sigma^\prime) = 0$. Then $\sigma$ coincides with $\sigma^\prime$ that all variables in $\lnot A_1 \lor \ldots \lor \lnot A_n$ are substituted with constants.
\end{proof}

\begin{lemma}
\label{lem:tres_guard}
In the application of \textbf{TRes}, resolvents of a loosely guarded clause and a set of (loosely) guarded clauses are (loosely) guarded clauses.
\end{lemma}
\begin{proof}
Let a flat loosely guarded clause $C = \lnot A_1 \lor \ldots \lor \lnot A_m \lor \ldots \lor \lnot A_n \lor D$ be the main premise and (loosely) guarded clauses $C_1 = B_1 \lor D_1, \ldots, C_n = B_n \lor D_n$ be the side premises in \textbf{TRes}, producing a resolvent $C^\prime = (D_1 \lor \ldots \lor D_m \lor \lnot A_{m+1} \lor \ldots \lor \lnot A_{n} \lor D)\sigma$, where $\sigma$ is an mgu among eligible literals $A_1, \ldots, A_m$ and $B_1, \ldots, B_m$. W.l.o.g. let $x$ be a top variable in $A_1$. \textsc{Lemma \ref{lem:tres_matching}} shows that $\sigma$ substitutes $x$ with either a constant or a compound term. 

Suppose $\sigma$ substitutes $x$ with a constant. Item 3 in \textsc{Lemma \ref{lem:tres_matching}} shows that $\sigma$ is a ground substitution that substitutes variables with constants. \textsc{Lemma \ref{lem:eligible_covering}} shows that an eligible literal contains all variables of that premise. Then all variables in \textbf{TRes} premises are substituted by constants. Hence the \textbf{TRes} resolvent is ground.

Suppose $\sigma$ substitutes $x$ with a compound term. $C_1$ is covering implies $\Var(x\sigma) = \Var(B_1\sigma)$, hence $\Var(x\sigma) = \Var(A_1\sigma)$. Because of the variable co-occurrence property (Condition 3 in \textsc{Definition \ref{def:gc}}) of loosely guarded clauses, $x$ occurs in each literal in $\lnot A_1 \lor \ldots \lor \lnot A_m$, thus $\Var(x\sigma) = \Var(A_1\sigma) = \ldots = \Var(A_m\sigma)$. Then $\Var(x\sigma) = \Var(A_1\sigma) = \ldots = \Var(A_m\sigma) = \Var(B_1\sigma) = \ldots = \Var(B_m\sigma)$. Let $C_1$ contain a guard $G$ (or loose guards $\mathcal{G}$), and literal $L$ and compound term $t$ occurs in either $C$, or a literal $C_1 \ldots, C_m$. \textsc{Lemma~\ref{lem:tres_matching}} shows that non-top variables in $A_1, \ldots, A_m$ match constants or variable terms in $B_1, \ldots, B_m$. Then variables in $B_1, \ldots, B_m$ can only be unified with either constants or variables. Then $G\sigma$ and $\mathcal{G}\sigma$ are flat. Since $\Var(\mathcal{G}\sigma) = \Var(G\sigma) = \Var(B_1\sigma)$ and $\Var(B_1\sigma) = \ldots = \Var(B_m\sigma) = \Var((A_1, \ldots, A_m)\sigma)$, according to \textsc{Lemma \ref{lem:eligible_covering}}, $\Var(\mathcal{G}\sigma) = \Var(G\sigma) = \Var(C_1\sigma) = \ldots = \Var(C_m\sigma) = \Var(C\sigma)$. Then using Item 3 in \textsc{Lemma \ref{lem:conclusion}}, $\Var(L\sigma) \subseteq \Var(A_i\sigma)$ or $\Var(L\sigma) \subseteq \Var(B_i\sigma)$, $\Var(L\sigma) \subseteq \Var(G\sigma)$ and $\Var(L\sigma) \subseteq \Var(\mathcal{G}\sigma)$. $G\sigma$ is a guard ($\mathcal{G}\sigma$ are guards) in $C^\prime$. \textsc{Lemma \ref{lem:tres_matching}} shows that $A_i\sigma$ and $B_i\sigma$ are simple, then using Item 1 in \textsc{Lemma \ref{lem:conclusion}}, since $\Var(L) \subseteq \Var(A_i)$ or $\Var(L) \subseteq \Var(B_i)$, $L\sigma$ is simple. Hence $C^\prime$ is simple. Using Item 5 in \textsc{Lemma \ref{lem:conclusion}}, since $\Var(t) = \Var(A_i)$ or $\Var(t) = \Var(B_i)$, $\Var(t\sigma) = \Var(A_i\sigma)$ or $\Var(t\sigma) = \Var(B_i\sigma)$, $\Var(t\sigma) = \Var(G\sigma) = \Var(\mathcal{G}\sigma)$. $C^\prime$ is covering. The above discussion immediately holds if $\sigma$ substitutes $x$ with a ground compound term. Therefore $C^\prime$ is a (loosely) guarded clause. 
\end{proof}

\begin{lemma}
\label{lem:bounded_width}
In applications of \textbf{Conden}, \textbf{Fact}, \textbf{Res} and \textbf{TRes} to (loosely) guarded clauses, let premises $C_1, \ldots, C_n$ derive a conclusion $C^\prime$. Then there is at least one $C_i$ in $C_1, \ldots, C_n$ satisfying that \text{$\Var(C^\prime) \subseteq \Var(C_i)$} up to variable renaming. 
\end{lemma}
\begin{proof}
In the applications of \textbf{Conden} and \textbf{Fact}, the (loosely) guards in $C^\prime$ are inherited from the single premise $C$, and in the applications of \textbf{Res} and \textbf{TRes}, the (loosely) guards in $C^\prime$ are inherited from the positive premise $C$. After variable renaming, (loosely) guards $\mathcal{L}$ in the conclusion cannot contain more types of variables than $\mathcal{L}$ in the premise, therefore $\Var(C^\prime) \subseteq \Var(C)$.
\end{proof}

\begin{example}
\label{example:unLGC}
Consider an unsatisfiable set of loosely guarded clauses $C_1, \ldots, C_9$:
\begin{gather*}
C_1 = \lnot A_1xy \lor \lnot A_2yz \lor \lnot A_3zx \lor Bxyb \qquad C_2 = A_3(x,fx) \lor \lnot G_3x \qquad C_3 = A_2(fx,fx) \lor \lnot G_2x \\
C_4 = A_1(fx,x) \lor D(gx) \lor \lnot G_1x \quad \ C_5 = \lnot Bxyb \quad \ C_6 = \lnot Dx \quad \ C_7 = G_1(fa) \quad \ C_8 = G_3(fa) \quad \ C_9 = G_2a
\end{gather*}
Suppose the precedence on which $\succ_{lpo}$ is based is $f > g > a > b > B > A_1 > A_2 > A_3 > D > G_1 > G_2 > G_3$ ($a$ and $b$ are constants). We annotate the $\succ_{lpo}$-maximal literals with "stars" as in $L^\ast$ and "box" the selected literals as in $\boxed{L}$. Then $C_1, \ldots, C_9$ are presented as:
\begin{gather*}
C_1 = \boxed{\lnot A_1xy} \lor \boxed{\lnot A_2yz} \lor \boxed{\lnot A_3zx} \lor Bxyb \qquad 
C_2 = A_3(x,fx)^\ast \lor \lnot G_3x \qquad C_3 = A_2(fx,fx)^\ast \lor \lnot G_2x \\
C_4 = A_1(fx,x)^\ast \lor D(gx) \lor \lnot G_1x \ \ C_5 = \boxed{\lnot Bxyb} \ \ C_6 = \boxed{\lnot Dx} \ \ 
C_7 = G_1(fa)^\ast \ \ C_8 = G_3(fa)^\ast \ \ C_9 = G_2a^\ast
\end{gather*}
One can use any clause to start the derivation, w.l.o.g., we start with $C_1$. For each newly derived clause, we immediately apply \textbf{Algorithms \ref{algorithm:refine}} to determine the eligible literals. 
\begin{enumerate}
\item $C_1$ satisfies the conditions in $\SelectT$, then one first selects all negative literals in $C_1$. If side premises of $C_1$ exist, then $\ComputeTop$ selects top-variable literals in $C_1$ . 
\item $C_2, C_3, C_4$ are found as $C_1$'s side premises. $\ComputeTop(C_2, C_3, C_4, C_1)$ computes an mgu $\sigma^\prime = \{x/ffx^\prime, y/fx^\prime, z/fx^\prime\}$, and $\SelectT$ selects $\lnot A_1xy$ and $\lnot A_3zx$ as $x$ is the only top variable in $C_1$.
\item \textbf{TRes} is applied to $C_1$, $C_2$ and $C_4$ with an mgu $\sigma = \{x/fx^\prime, y/x^\prime, z/x^\prime \}$, deriving $C_{10} = \lnot A_2xx \lor B(f(x),x,b)^\ast \lor D(gx) \lor \lnot G_1x \lor \lnot G_3x$ (after renaming $x^\prime$ with $x$). No resolution steps can be performed on $C_3$ and $C_{10}$ since they do not have complementary eligible literals, but an inference can be performed between $C_{5}$ and $C_{10}$. 
\item Applying \textbf{Res} to $C_{5}$ and $C_{10}$ derives $C_{11} = \lnot A_2xx \lor D(gx)^\ast \lor \lnot G_1x \lor \lnot G_3x$. 
\item Applying \textbf{Res} to $C_{6}$ and $C_{11}$ derives $C_{12} = \boxed{\lnot A_2xx} \lor \boxed{\lnot G_1x} \lor \boxed{\lnot G_3x}$.
\item Due to the presence of $C_3, C_7, C_8$ and $C_{12}$ satisfy the \textbf{TRes} conditions, $\ComputeTop(C_3, C_7, C_8, C_{12})$ finds the top variables in $C_{12}$, producing an mgu $\sigma^\prime = \{x/fa\}$. Then $x$ is the top variable in $C_{12}$, and $\SelectT(C_{12})$ selects all literals in $C_{12}$. Applying \textbf{TRes} to $C_3, C_7, C_8$ and $C_{12}$ derives $C_{13} = \boxed{\lnot G_2x}$. 
\item Applying \textbf{Res} to $C_9$ and $C_{13}$ derives $\bot$.
\end{enumerate}
Given a set of (loosely) guarded clauses, orderings and selection refinement allow inferences building a model or deriving a contradictory without producing unnecessary conclusions. \textsc{Example \ref{example:unLGC}} shows that by using the refinement \emph{T-Refine}, resolution computes fewer inferences to derive a contradiction. E.g., inferences between $C_2$ and $C_8$ and inferences between $C_3$ and $C_9$ are prevented since theses pairs do not contain complementary eligible literals. These unnecessary inferences would be computed if there is no refinement guiding resolution.
\end{example}

\section{Section \ref{sec:handle_query}: Handling query clauses}
\label{lem:handle_q}

\begin{lemma}
\label{lem:con_split}
In applications of \textbf{Conden} and \textbf{Split}, a query clauses $Q$ is replaced by a set of query clauses $Q_1, \ldots, Q_n$ where each $Q_i$ in $Q_1, \ldots, Q_n$ satisfies that $\Len(Q_i) < \Len(Q)$.
\end{lemma}
\begin{proof}
Follows from the definitions of \textbf{Conden} and \textbf{Split}.
\end{proof}

\begin{lemma}
\label{lem:sep_conlusion}
If \textbf{Sep} is applicable to a query clause, then it derives a query clause and a guarded clause.
\end{lemma}
\begin{proof}
Recall \textbf{Sep}
\begin{displaymath}
 \prftree[r,l]{\quad 
 \mbox{\vbox{\noindent
 if i) $A$ contains both isolated variables and chained variables, \\ ii) $\overline x = \Var(A) \cap \Var(D)$, iii) $\Var(C) \subseteq \Var(A)$, iv) $d_s$ is a fresh \\ predicate symbol.}}}{\textbf{Sep}: \ }
  {N \cup \{C \lor A \lor D\}}
  {N \cup \{C \lor A \lor d_s(\overline x), \lnot d_s(\overline x) \lor D\}}
\end{displaymath}
Suppose $C \lor A \lor D$ is a query clause. Then all literals in $C \lor A \lor D$ are negative and $C \lor A \lor D$ contains only variables and constants as arguments. Since $\Var(C) \subseteq \Var(A)$ (Condition iii) in \textbf{Sep}), $\overline x \in \Var(A)$ (Condition ii) in \textbf{Sep}) and $A$ is a negative literal, $A$ is a guard in $C \lor A \lor d_s(\overline x)$. Since $C \lor A \lor d_s(\overline x)$ does not contain compound term, covering property does not need to be checked. Thus $C \lor A \lor d_s(\overline x)$ is a guarded clause. Similarly, since $\lnot d_s(\overline x) \lor D$ is composed of negative literals and only variables and constants are arguments, it is a query clause. 
\end{proof}

\begin{lemma}
\label{lem:isolated_only}
An indecomposable query clause is an isolated-only query clause iff it is a guarded clause.
\end{lemma}
\begin{proof}
If: Let a query clause $Q$ be a guarded clause. Then surface literals $\mathcal{L}$ in $Q$ share a same variable set and each literal $L$ in $\mathcal{L}$ satisfies $\Var(Q) = \Var(L)$. This implies that there is no chained variable in $Q$, so that $Q$ is an isolated-only query clause.

Only if: Using the definition of chained variable, if a query clause $Q$ contains no chained variable, then either $Q$ contains only one surface literal, or all surface literals in $Q$ share a same variable set. Then for any surface literal $L$ in $Q$, $\Var(L) = \Var(Q)$. $Q$ is compound-term free, therefore $Q$ is a guarded clause.
\end{proof}

\begin{lemma}
\label{lem:GYO}
Exhaustively applying \textbf{Sep} to an indecomposable query clause $Q$ transforms it into:
\begin{enumerate}
\item a set of Horn guarded clauses if $Q$ is acyclic, or
\item a set of Horn guarded clauses and a \emph{chained-only query clause} if $Q$ is cyclic.
\end{enumerate}
\end{lemma}
\begin{proof}
By the definition of GYO reduction \cite{yu1979tree}.	
\end{proof}

\begin{lemma}
\label{lem:tres_on_chain}
In an application of \textbf{TRes} to a chained-only query clause $\lnot A_1 \lor \ldots \lor \lnot A_n$ and (loosely) guarded clauses $B_1 \lor D_1, \ldots, B_n \lor D_n$ where $\sigma$ is an mgu such that $B_i\sigma = A_i\sigma$ for all $i$ such that $1 \leq i \leq m$. Let $\{i_1, \ldots, i_k\}$ and $\{j_1, \ldots, j_h\}$ be two subsets of $\{1, \ldots, n\}$, and let these conditions hold:
\begin{itemize}
\item no top variable occurs in any literal in $\{A_{j_1}, \ldots, A_{j_h}\}$
\item $\mathcal{X}$ is a closed top variable set s.t. each literal in $\{A_{i_1}, \ldots, A_{i_k}\}$ contains at least one variable in $\mathcal{X}$.
\end{itemize}
Then 
\begin{enumerate}
\item $(\lnot A_{j_1} \lor \ldots \lor \lnot A_{j_h})\sigma$ is a query clause,
\item $(D_{i_1} \lor \ldots \lor D_{i_k})\sigma$ is a (loosely) guarded clause.
\end{enumerate}
\end{lemma}
\begin{proof}
Applying \textbf{TRes} to a chained-only query clause and (loosely) guarded clauses mean the conditions in \textsc{Lemma \ref{lem:tres_matching}} are satisfied. Item 2 in \textsc{Lemma \ref{lem:tres_matching}} shows that $\sigma$ substitutes non-top variables with either constants or variables. $\lnot A_{j_1} \lor \ldots \lor \lnot A_{j_h}$ contains only constants and variables as arguments without top variable occurring implies that $(\lnot A_{j_1} \lor \ldots \lor \lnot A_{j_h})\sigma$ contains only constants and variables as arguments. Hence $(\lnot A_{j_1} \lor \ldots \lor \lnot A_{j_h})\sigma$ is a query clause.

Item 2: Let a variable $x$ in $\mathcal{X}$ occur in $A_{i_1}$. Using Item 1 in \textsc{Lemma \ref{lem:tres_matching}}, $x$ matches either a constant or a compound term. Suppose $x$ matches a constant. Then Item 3 in \textsc{Lemma \ref{lem:tres_matching}} implies that the \textbf{TRes} resolvent is a ground clause, therefore $(D_{i_1} \lor \ldots \lor D_{i_k})\sigma$ is a (loosely) guarded clause. Now suppose $x$ matches a compound term in $B_{i_1}$. We show $D_{i_1}\sigma$ is a (loosely) guarded clause, and then show that $(D_{i_1} \lor \ldots \lor D_{i_k})\sigma$ is a (loosely) guarded clause. Suppose $G$ ($\mathcal{G}$) are (loose) guards, $t$ is a compound term, and $L$ is a literal in $D_{i_1}$. Then $\Var(G) = \Var(\mathcal{G}) = \Var(t)$ and $\Var(L) \subseteq \Var(G)$ ($\Var(L) \subseteq \Var(\mathcal{G})$). Item 1 in \textsc{Lemma \ref{lem:tres_matching}} implies that variable arguments in $B_{i_1}$ match only constant or variables. Hence $\sigma$ substitutes variables in~$B_{i_1}$ with either constant or variables. Then $G\sigma$ ($\mathcal{G}\sigma$) is flat and $\Var(G\sigma) = \Var(\mathcal{G}\sigma) = \Var(D_{i_1}\sigma)$, hence $G$ ($\mathcal{G}$) are the (loose) guards in $D_{i_1}\sigma$. Item 2 in \textsc{Lemma \ref{lem:conclusion}} implies $\Var(L\sigma) \subseteq \Var(G\sigma)$ ($\Var(L\sigma) \subseteq \Var(\mathcal{G})$), and Item 3 in \textsc{Lemma \ref{lem:conclusion}} implies $\Var(t\sigma) = \Var(G\sigma) = \Var(\mathcal{G}\sigma)$. Hence $D_{i_1}\sigma$ is a (loosely) guarded clause, therefore each clause in $D_{i_1}\sigma, \ldots, D_{i_k}\sigma$ is a (loosely) guarded clause. Suppose $D_{i_a}$ and $D_{i_b}$ are two clauses in $D_{i_1}, \ldots, D_{i_k}$. Let $x_a$ be a top variable in $A_{i_a}$ and $x_b$ be a top variable in $A_{i_b}$. Then $x_a, x_b \in \mathcal{X}$. Since $x_a$ and $x_b$ are connected, we can find a sequence of top variables $x_a, \ldots, x_b$ in $\mathcal{X}$ such that each pair of adjacent variables occurs in a literal in $\{\lnot A_{i_1}, \ldots, \lnot A_{i_k}\}$. Item 2 in \textsc{Lemma \ref{lem:tres_matching}} shows that each variable in $x_a, \ldots, x_b$ is substituted by a compound term. Then by the covering property, for every adjacent variables $x$ and $x^\prime$ in $x_a, \ldots, x_b$, $\Var(x\sigma) = \Var(x^\prime\sigma)$ (since they occur in the same literal). Then $\Var(x_a\sigma) = \Var(x_b\sigma)$. Suppose $x_a$ matches $t$ in $B_{i_a}$ and $x_b$ matches $s$ in $B_{i_b}$. The covering property means $\Var(t) = \Var(D_{i_a})$ and $\Var(s) = \Var(D_{i_b})$. Hence $\Var(t\sigma) = \Var(D_{i_a}\sigma)$ and $\Var(s\sigma) = \Var(D_{i_b}\sigma)$. $t$ matches $x_a$ and $s$ matches $x_b$, therefore $\Var(t\sigma) = \Var(x_a\sigma)$ and $\Var(s\sigma) = \Var(x_b\sigma)$. Then $\Var(x_a\sigma) = \Var(D_{i_a}\sigma)$ and $\Var(x_b\sigma) = \Var(D_{i_b}\sigma)$. Hence $\Var(D_{i_a}\sigma) = \Var(D_{i_b}\sigma)$. This means that clauses in $D_{i_1}\sigma, \ldots, D_{i_k}\sigma$ share the same variable set. Each clause in $D_{i_1}\sigma, \ldots, D_{i_k}\sigma$ is a (loosely) guarded clause and $\Var(D_{i_1}\sigma) = \ldots = \Var(D_{i_k}\sigma)$, therefore $(D_{i_1} \lor \ldots \lor D_{i_k})\sigma$ is a (loosely) guarded clause.
\end{proof}

\begin{lemma}
\label{lem:ttrans_conclusion}
Let \textbf{TRes} derive the resolvent $R$ using a chained-only query clause $Q$ and a set of (loosely) guarded clauses. Then \textbf{T-Trans} transforms $R$ into a set of (loosely) guarded clauses and a query clause $Q^\prime$ satisfying $\Len(Q^\prime) < \Len(Q)$.
\end{lemma}
\begin{proof}
Let \textbf{TRes} derive the resolvent $R = (\lnot A_{m+1} \lor \ldots \lor \lnot A_n \lor D_1 \lor \ldots \lor D_m)\sigma$ using (loosely) guarded clauses $B_1 \lor D_1, \ldots, B_n \lor D_n$ as the side premises, a chained-only query clause $Q = \lnot A_1 \lor \ldots \lor \lnot A_n$ as the main premise and a substitution $\sigma$ such that $B_i\sigma = A_i\sigma$ for all $i$ such that $1 \leq i \leq m$ as an mgu. Suppose $\mathcal{X}_1, \ldots, \mathcal{X}_t$ be all closed top variable sets and $\mathcal{A}_i$ is a subset of $\{A_1, \ldots, A_m\}$ such that each literal in $\mathcal{A}_i$ contains at least one variable in $\mathcal{X}_i$. Therefore we have top-variable literal sets $\mathcal{A}_1, \ldots, \mathcal{A}_t$ such that $\mathcal{A}_1 \cup \ldots \cup \mathcal{A}_t$ is equivalent to $\{A_1, \ldots, A_m\}$. For each pair $\mathcal{A}_i$ and $\mathcal{A}_j$ in $\mathcal{A}_1, \ldots, \mathcal{A}_t$, $\mathcal{A}_i$ and $\mathcal{A}_j$ are top variable disjoint since overlapping top variables makes $\mathcal{X}_i$ and $\mathcal{X}_j$ being a single closed top variable set. Since $\mathcal{A}_1, \ldots, \mathcal{A}_t$ is a partition of $\{A_1, \ldots, A_m\}$, we can present $(\lnot A_{m+1} \lor \ldots \lor \lnot A_n \lor D_1 \lor \ldots \lor D_m)\sigma$ as $(\lnot A_{m+1} \lor \ldots \lor \lnot A_n \lor \mathcal{D}_1 \lor \ldots \lor \mathcal{D}_t)\sigma$ where $\mathcal{D}_i$ presents $D_a \lor \ldots \lor D_b$ whenever $\mathcal{A}_i$ presents $A_a \lor \ldots \lor A_b$. \textsc{Lemma \ref{lem:tres_on_chain}} shows that each $\mathcal{D}_i\sigma$ is a (loosely) guarded clause. Adding positive compound-term free literals $L$, satisfying that $\Var(\mathcal{D}_i\sigma) = \Var(L)$, to $\mathcal{D}_i\sigma$ makes $\mathcal{D}_i\sigma \lor L$ a (loosely) guarded clause. Adding negative compound-term free literal $\lnot L$ to a query clause $Q$ makes $Q \lor \lnot L$ a query clause. Then \textbf{T-Trans} transforms $R$ into a query clause $Q^\prime = (\lnot A_{m+1} \lor \ldots \lor \lnot A_n)\sigma \lor \lnot d_s^1 \lor \ldots \lor \lnot d_s^t$ and (loosely) guarded clauses $d_s^1 \lor \mathcal{D}_1\sigma, \ldots, d_s^t \lor \mathcal{D}_t\sigma$.

Each $\mathcal{X}_i$ contains at least one top variable, namely $x$. Since $x$ is a chained variable, $x$ occurs at least in two literals in $\{A_1, \ldots, A_m\}$, therefore $\mathcal{A}_i$ contains at least two literals, namely $A_i$ and $A_j$. Then $\mathcal{D}_i$ presents $D_i$ and $D_j$. We can write $R$ as $(\lnot A_{m+1} \lor \ldots \lor \lnot A_n \lor D_i \lor D_j \ldots)\sigma$. Then \textbf{T-Trans} introduces one definer for at least $D_i$ and $D_j$ ($\mathcal{D}_i\sigma$), to obtain $Q^\prime$. Hence $Q^\prime$ contains a smaller number of literals than $Q$.
\end{proof}

\begin{lemma}
\label{lem:smaller_query}
In applications of \textbf{Sep} or a combination of \textbf{T-TRes} and \textbf{T-Trans}, a conclusion is smaller than any of its premises w.r.t. $\succ_{lpo}$.
\end{lemma}
\begin{proof}
In an application of \textbf{Sep} or a combination of \textbf{T-TRes} and \textbf{T-Trans}, an introduced definers in the conclusion is smaller than any other predicates symbols in the premises. Then the lemma holds. 
\end{proof}

\begin{example}
\label{example:multicycles}
Consider the guarded clauses
\begin{gather*}
 C_1 = A_1(fxy,fxy) \lor D_1(h_1xy) \lor \lnot G_1xy \qquad C_2 = A_2(fxy,x) \lor \lnot G_2xy \qquad C_3 =  A_3(fxy,x) \lor \lnot G_3xy \\
 C_4 = A_4(x, fxz) \lor \lnot G_4xz \qquad C_5 = A_5(x, fxz) \lor \lnot G_5xz \qquad C_6 = A_6(fxz, fxz) \lor D_2(h_2xz) \lor \lnot G_6xz \\  C_7 = B(gx) \lor G_7x 
\end{gather*}
and the chain-only query clause from \emph{Fig. \ref{fig:two_cycles}}, i.e.,
\begin{gather*}
Q = \lnot A_1(x_1, x_2) \lor \lnot A_2(x_1, x_3) \lor \lnot A_3(x_2, x_3) \lor \lnot A_4(x_3, x_4) \lor \lnot A_5(x_3, x_5) \lor \lnot A_6(x_4, x_5) \lor \lnot B(x_3)
\end{gather*}
$\ComputeTop$ computes the mgu $\sigma = \{x_1/f(gx, y), x_2/f(gx, y), x_3/gx, x_4/f(gx, z), x_5/f(gx, z)\}$ among $Q$ and $C_1, \ldots, C_7$, therefore $x_1, x_2, x_4, x_5$ are top variables. Since these variables occurs in $A_1, \ldots, A_6$ in $Q$, $\SelectT$ selects $A_1, \ldots, A_6$ in $Q$. \textbf{TRes} is performed on $Q$ and $C_1, \ldots, C_6$. The resolvent $R = D_1(h_1xy) \lor \lnot G_1xy \lor \lnot G_2xy \lor \lnot G_3xy \lor D_2(h_2xz) \lor \lnot G_6xz \lor \lnot G_4xz \lor \lnot G_5xz \lor \lnot Bx$. Notice that $R$ consists of two guarded clauses, namely $D_1(h_1xy) \lor \lnot G_1xy \lor \lnot G_2xy \lor \lnot G_3xy$ and $D_2(h_2xz) \lor \lnot G_6xz \lor \lnot G_4xz \lor \lnot G_5xz$, and the query clause $\lnot Bx$.

Now we explain how \textbf{Algorithm \ref{algorithm:closed_top}} finds closed top variable sets in $Q$ and how \textbf{T-Trans} identifies (loosely) guarded clauses in $R$. Let $\mathcal{X}$ be the set of top variables $\{x_1, x_2, x_4, x_5\}$ in $Q$. We pick a variable $x_1$ from $\mathcal{X}$, and use \textbf{Algorithm \ref{algorithm:closed_top}} to find $x_1$-connected top variables. $x_1$-containing literals in $Q$ are $A_1$ and $A_2$, therefore find the $x_1$-connected top variables $\{x_1, x_2\}$ in $A_1$ and $A_2$. Using \textbf{Algorithm \ref{algorithm:closed_top}} again with $\mathcal{X}$ and $x_2$, we find the $x_2$-containing literals $A_1$ and $A_3$, and realise that $A_3$ only contains the top variable $x_2$. No new $x_1$-connected top variables are found, hence \textbf{Algorithm \ref{algorithm:closed_top}} terminates. Then $\{x_1, x_2\}$ is a closed top variable set, and we remove $\{x_1, x_2\}$ from $\{x_1, x_2, x_4, x_5\}$, obtaining $\{x_4, x_5\}$. Similarly, we pick a variable from $\{x_4, x_5\}$ and apply \textbf{Algorithm \ref{algorithm:closed_top}} again. Eventually we find that $\{x_4, x_5\}$ is a closed top variable set. Therefore, $\{x_1, x_2, x_4, x_5\}$ is partitioned into two closed top variable sets: $\{x_1, x_2\}$ and $\{x_4, x_5\}$.

Using a closed top variable set $\{x_1, x_2\}$, \textbf{T-Trans} finds the remainders of side premises that map to $\{x_1, x_2\}$. $\{x_1, x_2\}$ occurs in the $A_1, A_2$ and $A_3$ literals in $Q$, which map to positive $A_1, A_2$ and $A_3$ literals in $C_1$, $C_2$ and $C_3$, respectively. Therefore \textbf{T-Trans} introduces a definer $\lnot d_t^1(x,y)$ for the remainders of $C_1$, $C_2$ and $C_3$ in $R$, i.e., $D_1(h_1xy) \lor \lnot G_1xy \lor \lnot G_2xy \lor \lnot G_3xy$. Similarly $\{x_4, x_5\}$ maps to $C_4$, $C_5$ and $C_6$, therefore \textbf{T-Trans} introduces a definer $d_t^2(x,z)$ for the remainders of $C_4$, $C_5$ and $C_6$ in $R$, i.e., $D_2(h_2xz) \lor \lnot G_6xz \lor \lnot G_4xz \lor \lnot G_5xz$. Therefore \textbf{T-Trans} transforms the \textbf{TRes} resolvent $R$ into 
\begin{itemize}
\item a guarded clause $D_1(h_1xy) \lor \lnot G_1xy \lor \lnot G_2xy \lor \lnot G_3xy \lor d_t^1(x,y)$, and
\item a guarded clause $D_2(h_2xz) \lor \lnot G_6xz \lor \lnot G_4xz \lor \lnot G_5xz \lor d_t^2(x,z)$, and
\item a query clause $Q^\prime = \lnot Bx \lor \lnot d_t^1(x,y) \lor \lnot d_t^2(x,z)$.
\end{itemize}
Note that $Q^\prime$ is smaller than $Q$. 
\end{example}

\section{Section \ref{sec:query_lgf}: Querying GF and LGF}
\begin{lemma}
\label{lem:clause_formula}
By unskolemising and negating a (loosely) guarded clause and a query clause, it is transformed into a first-order sentence.
\end{lemma}
\begin{proof}
Since $Q$ contains only constants and variables as arguments, thus the unskolemisation is unnecessary. Adding universal quantifications to $Q$ and negate the result, one obtains a BCQ.

Let a (loosely) guarded clause $C$ be a form of $A_1(f_1(\overline x), \ldots, f_n(\overline x), x_1, \ldots, x_a) \lor \lnot A_2(g_1(\overline x), \ldots, g_m(\overline x), x_1, \\ s\ldots, x_b) \lor B(x_1, \ldots, x_c) \lor \lnot D(x_1, \ldots, x_d) \lor \lnot G(\overline x)$ where $\lnot G(\overline x)$ is the (loosely) guard and $\Var(C) = \overline x$. $C$ is able to generalise all possible (loosely) guard clauses. Since $C$ is covering, one can unskolemise $C$ and write $C$ as a first-order formula $F = \forall \overline x (\lnot G(\overline x) \lor \exists \overline y A_1(\overline y, x_1, \ldots, x_a) \lor \exists \overline z \lnot A_2(\overline z, x_1, \ldots, x_b) \lor B(x_1, \ldots, x_c) \lor \lnot D(x_1, \ldots, x_d))$ (one can distribute existential quantifications over disjunction whenever literals share the same Skolem function). Since first-order logic is closed under negation, $\lnot F$ is also a closed first-order formula, thus a first-order sentence.
\end{proof}

\begin{lemma}
\label{lem:finite_definer}
In the application of \textbf{Q-AR} to solve BCQ answering and rewriting problem over the (loosely) guarded fragment, only finitely many definers are introduced.
\end{lemma}
\begin{proof}
Definers are introduced by either \textbf{Q-Trans}, or \textbf{Sep} or \textbf{T-Trans}. 

Before applying inference rules, \textbf{Q-Trans} introduces definers for each universally quantified subformula in a (loosely) guarded formula. Hence if the number of universally quantified formulas is $n$, \textbf{Q-Trans} introduces at most $n$ definers. \textbf{Sep} is used to preprocess query clauses $Q$, before any inference rule performs on $Q$, therefore $Q$ contains $n$ literals, \textbf{Sep} introduces at most $n$ definers. \textbf{T-Trans} is used together with \textbf{TRes}. Let $Q$ be the query clause in the main premise of \textbf{TRes}, and produce the resolvent $Q$, and $Q^\prime$ be the query clause produced by applying \textbf{T-Trans} to $R$. Then \textsc{lemma \ref{lem:ttrans_conclusion}} shows $\Len(Q^\prime) < \Len(Q)$. This means each time \textbf{T-Trans} introduces a definer for a query clause $Q$, the length of new query clause is smaller than $Q$. Hence at most $n$ definers are need when applying \textbf{Q-TRes} and \textbf{Q-Trans} to an $n$-length query clause.  
\end{proof}

\begin{theorem}
\label{thm:termination}
In the application of \textbf{Q-AR} to solve BCQ answering and rewriting problem over the (loosely) guarded fragment, \textbf{Q-AR} terminates in finitely bounded time. 
\end{theorem}
\begin{proof}
By \textsc{theorem \ref{thm:decide}}, \textsc{lemma \ref{lem:smaller_query}} and \textsc{lemma \ref{lem:finite_definer}}.
\end{proof}

\begin{theorem}[Soundness of \textbf{Q-AR}]
\label{thm:sound}
\textbf{Q-AR} is a sound inference system.
\end{theorem}
\begin{proof}
By \textsc{theorem \ref{thm:tinf_sound}} and the fact that renaming \textbf{T-Trans} preserves satisfiability.
\end{proof}

\begin{theorem}
\label{thm:refu_complete}
Let $N$ be a set of clauses that is saturated up to redundancy with respect to \textbf{Q-AR}. Then $N$ is unsatisfiable if and only if $N$ contains an empty clause.
\end{theorem}
\begin{proof}
By \textsc{lemma \ref{lem:finite_definer}} and \textsc{theorem \ref{thm:tinf_refu}}.
\end{proof}

\bibliographystyle{siamplain}
\bibliography{reference.bib}

\begin{thebibliography}{10}

\bibitem{abiteboul1995chase}
{\sc S.~Abiteboul, R.~Hull, and V.~Vianu}, {\em Foundations of Databases: The
  Logical Level}, Addison-Wesley Longman Publishing Co., Inc., 1995.

\bibitem{andrka1998bounded}
{\sc H.~Andr{\'e}ka, I.~N{\'e}meti, and J.~van Benthem}, {\em Modal languages
  and bounded fragments of predicate logic}, J. Philos. Logic, 27 (1998),
  pp.~217--274.

\bibitem{bachmair1990restrictions}
{\sc L.~Bachmair and H.~Ganzinger}, {\em On restrictions of ordered
  paramodulation with simplification}, in Proc. CADE'90, vol.~449 of LNCS,
  Springer, 1990, pp.~427--441.

\bibitem{bachmair1997theory}
{\sc L.~Bachmair and H.~Ganzinger}, {\em A theory of resolution}, Research
  Report MPI-I-97-2-005, 1997.

\bibitem{bachmair2001resolution}
{\sc L.~Bachmair and H.~Ganzinger}, {\em Resolution theorem proving}, in
  Handbook of Automated Reasoning, A.~Robinson and A.~Voronkov, eds., Elsevier
  and {MIT} Press, 2001, pp.~19--99.

\bibitem{bachmair1993superposition}
{\sc L.~Bachmair, H.~Ganzinger, and U.~Waldmann}, {\em Superposition with
  simplification as a decision procedure for the monadic class with equality},
  in In Proc. KGC'93, vol.~713 of LNCS, Springer, 1993, pp.~83--96.

\bibitem{baget2011decidability}
{\sc J.-F. Baget, M.~Lecl{\' e}re, M.-L. Mugnier, and E.~Salvat}, {\em On rules
  with existential variables: Walking the decidability line}, Artif. Int., 175
  (2011), pp.~1620--1654.

\bibitem{barany2010querying}
{\sc V.~B{\'a}r{\'a}ny, G.~Gottlob, and M.~Otto}, {\em Querying the guarded
  fragment}, in Proc. LICS'10, IEEE, 2010, pp.~1--10.

\bibitem{barany2015negation}
{\sc V.~B\'{a}r\'{a}ny, B.~ten Cate, and L.~Segoufin}, {\em Guarded negation},
  J. ACM, 62 (2015), pp.~22:1--22:26.

\bibitem{calautti2015chase}
{\sc M.~Calautti, G.~Gottlob, and A.~Pieris}, {\em Chase termination for
  guarded existential rules}, in Proc. PODS'15, ACM, 2015, pp.~91--103.

\bibitem{cali2013taming}
{\sc A.~Cal\`{\i}, G.~Gottlob, and M.~Kifer}, {\em Taming the infinite chase:
  Query answering under expressive relational constraints}, J. Artif. Int.
  Res., 48 (2013), pp.~115--174.

\bibitem{calvanese2007tractable}
{\sc D.~Calvanese, G.~De~Giacomo, D.~Lembo, M.~Lenzerini, and R.~Rosati}, {\em
  Tractable reasoning and efficient query answering in description logics: The
  {DL-Lite} family}, J. Automat. Reasoning, 39 (2007), pp.~385--429.

\bibitem{chandra1977querycontainment}
{\sc A.~K. Chandra and P.~M. Merlin}, {\em Optimal implementation of
  conjunctive queries in relational data bases}, in Proc. SToC'77, ACM, 1977,
  pp.~77--90.

\bibitem{de2003deciding}
{\sc H.~de~Nivelle and M.~de~Rijke}, {\em Deciding the guarded fragments by
  resolution}, J. Symb. Comput., 35 (2003), pp.~21--58.

\bibitem{dershowitz1982ordering}
{\sc N.~Dershowitz}, {\em Orderings for term-rewriting systems}, Theoretical
  Comp. Sci., 17 (1982), pp.~279--301.

\bibitem{fermuller2001decision}
{\sc C.~G. Ferm{\"{u}}ller, A.~Leitsch, U.~Hustadt, and T.~Tammet}, {\em
  Resolution decision procedures}, in Handbook of Automated Reasoning, J.~A.
  Robinson and A.~Voronkov, eds., Elsevier and {MIT} Press, 2001,
  pp.~1791--1849.

\bibitem{ganzinger1999superposition}
{\sc H.~Ganzinger and H.~de~Nivelle}, {\em A superposition decision procedure
  for the guarded fragment with equality}, in Proc. LICS'99, IEEE, 1999,
  pp.~295--303.

\bibitem{ganzinger1998resolution}
{\sc H.~Ganzinger, U.~Hustadt, C.~Meyer, and R.~A. Schmidt}, {\em A
  resolution-based decision procedure for extensions of {K}4}, in Proc.
  {A}i{ML}'98, CSLI, 1998, pp.~225--246.

\bibitem{dixon1998resolution}
{\sc C.~Geissler and K.~Konolige}, {\em A resolution method for quantified
  modal logics of knowledge and belief}, in Proc. TARK'86, Morgan Kaufmann,
  1986, pp.~309--324.

\bibitem{gottlob2003hypertree}
{\sc G.~Gottlob, N.~Leone, and F.~Scarcello}, {\em Robbers, marshals, and
  guards: Game theoretic and logical characterizations of hypertree width}, J.
  Comp. and Syst. Sci., 66 (2003), pp.~775--808.

\bibitem{gradel1999guarded}
{\sc E.~Gr{\"a}del}, {\em Decision procedures for guarded logics}, in Proc.
  CADE'16, vol.~1632 of LNCS, Springer, 1999, pp.~31--51.

\bibitem{erich1999guards}
{\sc E.~Gr{\"a}del}, {\em On the restraining power of guards}, J. Symb. Logic,
  64 (1999), pp.~1719--1742.

\bibitem{GSS2001cq}
{\sc M.~Grohe, T.~Schwentick, and L.~Segoufin}, {\em When is the evaluation of
  conjunctive queries tractable?}, in Proc. STOC ’01, ACM, 2001,
  p.~657–666.

\bibitem{hirsch2002tableau}
{\sc C.~Hirsch and S.~Tobies}, {\em A tableau algorithm for the clique guarded
  fragment}, in Proc. AiML'00, World Scientific, 2000, pp.~257--277.

\bibitem{hladik2002saga}
{\sc J.~Hladik}, {\em Implementation and optimisation of a tableau algorithm
  for the guarded fragment}, in Proc. TABLEAUX'02, vol.~2381 of LNCS, Springer,
  2002, pp.~145--159.

\bibitem{hadkinson2002loosely}
{\sc I.~Hodkinson}, {\em Loosely guarded fragment of first-order logic has the
  finite model property}, Studia Logica, 70 (2002), pp.~205--240.

\bibitem{hustadt1999resolution}
{\sc U.~Hustadt}, {\em Resolution Based Decision Procedures for Subclasses of
  First-order Logic}, PhD thesis, Universit{\"a}t des Saarlandes,
  Saarbr{\"{u}}cken, Germany, 1999.

\bibitem{hustadt1997evaluating}
{\sc U.~Hustadt and R.~A. Schmidt}, {\em On evaluating decision procedures for
  modal logic}, in Proc. IJCAI'97, Morgan Kaufmann, 1997, pp.~202--207.

\bibitem{hustadt1999maslov}
{\sc U.~Hustadt and R.~A. Schmidt}, {\em Maslov's class {K} revisited}, in
  Proc. CADE'99, vol.~1632 of LNCS, Springer, 1999, pp.~172--186.

\bibitem{kazakov2006Phd}
{\sc Y.~Kazakov}, {\em Saturation-Based Decision Procedures for Extensions of
  the Guarded Fragment}, PhD thesis, Universit{\"a}t des Saarlandes,
  Saarbr{\"u}cken, Germany, 2006.

\bibitem{kikot2012conjunctive}
{\sc S.~Kikot, R.~Kontchakov, and M.~Zakharyaschev}, {\em Conjunctive query
  answering with {OWL 2 QL}}, in Proc. KR'12, AAAI, 2012, pp.~275--285.

\bibitem{max2013cq}
{\sc D.~Marx}, {\em Tractable hypergraph properties for constraint satisfaction
  and conjunctive queries}, J. ACM, 60 (2013).

\bibitem{mora2014kyrie2}
{\sc J.~Mora, R.~Rosati, and O.~Corcho}, {\em Kyrie2: Query rewriting under
  extensional constraints in {$\mathcal{ELHOI}$}}, in Proc. ISWC'14, vol.~8796
  of LNCS, Springer, 2014, pp.~568--583.

\bibitem{nonnengart2001computing}
{\sc A.~Nonnengart and C.~Weidenbach}, {\em Computing small clause normal
  forms}, in Handbook of Automated Reasoning, A.~Robinson and A.~Voronkov,
  eds., Elsevier and {MIT} Press, 2001, pp.~335--367.

\bibitem{rosati2010improving}
{\sc R.~Rosati and A.~Almatelli}, {\em Improving query answering over {DL-Lite}
  ontologies}, in Proc. KR'10, AAAI, 2010, pp.~290--300.

\bibitem{schmidt2000fluted}
{\sc R.~A. Schmidt and U.~Hustadt}, {\em A resolution decision procedure for
  fluted logic}, in Proc. CADE'00, vol.~1831 of LNCS, Springer, 2000,
  pp.~433--448.

\bibitem{van1997dynamic}
{\sc J.~van Benthem}, {\em Dynamic bits and pieces}, Research Report LP-97-01,
  Univ. Amsterdam, 1997.

\bibitem{vardi1996robust}
{\sc M.~Y. Vardi}, {\em Why is modal logic so robustly decidable?}, in Proc.
  {DIMACS} Workshop'96, {DIMACS/AMS}, 1996, pp.~149--183.

\bibitem{vardi2000constraint}
{\sc M.~Y. Vardi}, {\em Constraint satisfaction and database theory: A
  tutorial}, in Proc. PODS'00, ACM, 2000, pp.~76--85.

\bibitem{yannakakis1981acyclic}
{\sc M.~Yannakakis}, {\em Algorithms for acyclic database schemes}, in Proc.
  VLDB'81, VLDB Endowment, 1981, pp.~82--94.

\bibitem{yu1979tree}
{\sc C.~Yu and M.~Ozsoyoglu}, {\em An algorithm for tree-query membership of a
  distributed query}, in Proc. COMPSAC'79, IEEE, 1979, pp.~306--312.

\bibitem{zheng2020horn}
{\sc S.~Zheng and R.~A. Schmidt}, {\em Deciding the loosely guarded fragment
  and querying its {H}orn fragment using resolution}, in Proc. AAAI'20, AAAI,
  2020, pp.~3080--3087.

\end{thebibliography}
\end{document}